\documentclass[lettersize,journal]{IEEEtran}
\usepackage{amsmath,amsfonts}
\usepackage{algorithmic}
\usepackage{array}
\usepackage[caption=false,font=normalsize,labelfont=sf,textfont=sf]{subfig}
\usepackage{textcomp}
\usepackage{stfloats}
\usepackage{url}
\usepackage{verbatim}
\usepackage{graphicx}
\usepackage{soul}
\usepackage{tikz}
\usepackage{booktabs}
\usepackage{multirow}
\usepackage{amsmath}
\usepackage{amssymb}
\usepackage{tikz}
\usepackage{booktabs}
\usepackage{multirow}
\usepackage{cite}  % comment out for biblatex with backend=biber
\usepackage{array}
\usepackage{siunitx}  % For number/unit formatting (\num, \SI)
\usepackage{adjustbox}

\def\BibTeX{{\rm B\kern-.05em{\sc i\kern-.025em b}\kern-.08em
    T\kern-.1667em\lower.7ex\hbox{E}\kern-.125emX}}
\usepackage{balance}

\usepackage{placeins}
\newlength{\figvspace}
\usepackage{xcolor}
\usepackage{pict2e}

\newsavebox{\ORCIDlogo}
\savebox{\ORCIDlogo}{%
\setlength{\unitlength}{\dimexpr 1em/256\relax}%
\begin{picture}(256,256)%
  \color[HTML]{A6CE39}\put(128,128){\circle*{256}}%
  \color{white}%
  \put(78.6,199.2){\circle*{20}}%
  \moveto(70.9,176,9)\lineto(86.3,176,9)\lineto(86.3,69.8)\lineto(70.9,69.8)%
  \closepath\fillpath%
  \moveto(108.9,176.9)\lineto(150.5,176.9)%
  \curveto(190.1,176.9)(207.5,148.6)(207.5 ,123.3)%
  \curveto(207.5,95,8)(186,69.7)(150.7,69.7)%
  \lineto(108.9,69.7)%
  \closepath\fillpath%
  \color[HTML]{A6CE39}%
  \moveto(124.3,83.6)\lineto(148.8,83.6)%
  \curveto(183.7,83.6)(191.7,110.1)(191.7,123.3)%
  \curveto(191.7,144.8)(178,163)(148,163)%
  \lineto(124.3,163)%
  \closepath\fillpath%
\end{picture}%
}

\newcommand\orcidicon[1]{\href{https://orcid.org/#1}{\usebox{\ORCIDlogo}}}

\usepackage[hidelinks]{hyperref} %<--- Load after everything else

\usepackage{etoolbox}
\newbool{showR}
\setbool{showR}{false}

\definecolor{changed}{RGB}{0,0,0}

\begin{document}
\title{Glyph-Based Multiscale Visualization \\ of Turbulent Multi-Physics Statistics}

\author{\parbox{\textwidth}{\centering A. Cowe$^{1}$\orcidicon{0009-0004-5512-4227}, Tyson Neuroth$^{1}$\orcidicon{0000-0002-5866-5959}, Qi  Wu$^{2}$\orcidicon{0009-0004-5512-4227}, Martin Rieth$^{3}$\\\orcidicon{0000-0002-1967-6591}, Jacqueline Chen$^{3}$ 
\orcidicon{0000-0002-9268-0634}, Myoungkyu Lee$^{4}$
\orcidicon{0000-0002-5647-6265}, Kwan-Liu Ma$^{1}$\orcidicon{0000-0001-8086-0366}}
\\
% For Computer Graphics Forum: Please use the abbreviation of your first name.
{\parbox{\textwidth}{\centering $^1$ University of California, Davis, Davis CA, USA \\ $^2$ NVIDIA Research, Santa Clara CA, USA \\ $^3$ Sandia National Laboratories, Livermore CA, USA \\ $^4$ University of Houston, Houston TX, USA}}}

% \markboth{Journal of \LaTeX\ Class Files,~Vol.~18, No.~9, September~2020}%
% {How to Use the IEEEtran \LaTeX \ Templates}

\maketitle

\begin{abstract}
Many scientific and engineering problems involving multi-physics span a wide range of scales. Understanding the interactions across these scales is essential for fully comprehending such complex problems. However, visualizing multivariate, multiscale data within an integrated view where correlations across space, scales, and fields are easily perceived remains challenging. To address this, we introduce a novel local spatial statistical visualization of flow fields across multiple fields and turbulence scales. Our method leverages the curvelet transform for scale decomposition of fields of interest, a level-set-restricted centroidal Voronoi tessellation to partition the spatial domain into local regions for statistical aggregation, and a set of glyph designs that combines information across scales and fields into a single, or reduced set of perceivable visual representations. Each glyph represents data aggregated within a Voronoi region and is positioned at the Voronoi site for direct visualization in a 3D view centered around flow features of interest. We implement and integrate our method into an interactive visualization system where the glyph-based technique operates in tandem with linked 3D spatial views and 2D statistical views, supporting a holistic analysis. We demonstrate with case studies visualizing turbulent combustion data—multi-scalar compressible flows—and turbulent incompressible channel flow data. This new capability enables scientists to better understand the interactions between multiple fields and length scales in turbulent flows.
\end{abstract}

\begin{IEEEkeywords}
Multiscale decomposition,
glyphs,
isosurfaces,
statistical summarization,
scientific visualization.
\end{IEEEkeywords}

\section{Introduction} 
\label{sec:intro}

\IEEEPARstart{U}{nderstanding} turbulent flow physics is crucial in many scientific and engineering problems. For example, turbulence modulates reactions and transport in determining the efficiency and emissions from utilizing sustainable fuels, like hydrogen and ammonia, in engines for power and transportation. Geophysical turbulent flow affects the layout of turbines in wind farms to effectively harness wind energy. Turbulent transport and reactions behind shocks generated in high-speed flows determine the effectiveness of various flame holding strategies in hypersonic flight. Despite the governing equations of turbulent flow being well-known, the inherent multi-scale and nonlinear characteristics of turbulence coupled with multi-physics (e.g., chemical reactions, atomization, radiation) pose significant challenges to fundamental understanding and its control. While advances in high-performance computing have enabled the creation of rich datasets through direct numerical simulations (DNSs), providing accurate statistics to develop new models, the chaotic states and superposed structures at different length scales make studying instantaneous structures difficult. For instance, wall-bounded turbulence involves a wide spectrum of length scales whose importance varies with distance from the wall. In high-speed flows, the separation of these scales widens, affecting phenomena like wall shear stress and energy exchange between velocity components. Consequently, the superposition of these scales complicates understanding energy exchange mechanisms within coherent structures of turbulent flows. 
This complexity further increases when turbulent flows are coupled with multi-physics phenomena. In turbulent reacting flows, for example, the interplay between extraordinarily thin flames and turbulence across decades of length and time scales is central to developing more efficient and cleaner combustion systems. This complexity is heightened in hydrogen-containing, fuel-lean mixtures due to thermo-diffusive instabilities, where flames form intrinsic cellular patterns at small diffusive-reaction scales with length scales modulated by turbulence spanning a broad range of scales from the size of a combustor down to viscous scales where kinetic energy is dissipated. The intrinsic flame instability results in higher burning rates through preferential enrichment of the highly mobile hydrogen fuel depending upon the local flame topology. In turn, the gas expansion from the higher burning rate results in small-scale production of kinetic energy which nonlinearly interacts with kinetic energy generated by large-scale mean velocity gradients through the turbulence cascade and is eventually dissipated. Therefore, advanced analysis and visualization tools are needed to unravel these intricate multi-scale, multi-physics interactions.

From this, we identify two key steps. The first is to separate information by relevant scale and then correlate the emergent spectral characteristics with the spatial features of structures. The second is to visualize the multiscale multiphysics information obtained from the first. For the first step, there exist numerous Fourier-based methods for scale decomposition, including band-pass filtering with discrete Fourier transform and discrete wavelet transform. Of these, we utilize the discrete curvelet transform (CTF)~\cite{doi:10.1137/05064182X} for its preservation of geometric structures across scales, and its ability to relate global spectral information by scale to the local spatial orientation and position of corresponding signal sources via curvelets. It is also important that the scale decomposition preserves spectral energy. Parseval's theorem guarantees this property for %the DFT, orthogonal wavelets, and% 
the CTF stemming from its tight frame property. Thus, the CTF's properties make it well-suited for multiscale geometric analysis, as seen in the works of Bermejo et al.~\cite{BERMEJO-MORENO_PULLIN_2008, BUMANN2022105665}. %The second step is to visualize the multiscale multiphysics information obtained from the first.% 

The CTF scale decomposition yields additional data per scale and field, whose size matches that of the original data. Visualization is therefore made challenging by the large size and quantity of incoming data, which introduces additional storage demands and perceptual demands stemming from the sheer amount of information to be visually processed at once.
Thus, it is crucial that this information be visualized comprehensively and coherently. Spatial statistical aggregation is one effective approach for visualizing large multivariate volumetric data as it organizes it %the data
into smaller regions of aggregate statistics. In prior work on flow visualization, local statistics are computed based on topological features of interest~\cite{bremer2009topological}, then aggregated about volumetric flow features~\cite{9904439}. We advance this direction by applying the level set restricted centroidal Voronoi tessellation (LSRCVT)~\cite{9904439} to multivariate multiscale data, then extending the visualization to visually correlate these statistics with their spatial location in a 3D view. Using the level set restricted centroidal Voronoi tessellation (LSRCVT)~\cite{9904439}, data across all fields and scales is spatially segmented and organized such that it aligns with the flow features from a select field. In combustion analysis, we organize the statistics and glyphs around the flame surface defined by temperature $T$ in kelvin (K). In channel flow analysis, they are organized around coherent structures defined by turbulent kinetic energy $\epsilon_K$. 
With this segmentation, we aggregate the spectral energy derived from the scale decomposition by scale, field, and region (see Section~\ref{sec:aggregation}). This reduces the size of data needed to represent the multiscale multivariate information by several orders of magnitude, making its management and visualization more feasible. 

From here, the remaining challenge comes in visualizing the local multivariate-multiscale statistics. One approach might be to directly encode information across the aggregate region or surface with color seen in other similar approaches~\cite{ccd78f3da756466eb378417b08e49996, 7042344, :10.2312/VisSym/EuroVis07/011-018}. However, color alone may prove insufficient as it can only encode one field at a time. Additional color blending and intensity could be used to encode more information but this may obfuscate the multiscale information. Instead, we might consider using symbols alongside color to encode more information. This prompts the use of glyphs as they can present a lot of information in a singular visual representation. Since glyphs are commonly used to symbolize information at a particular point or region in space, they are a good fit for visualizing aggregate data across a spatial surface. Once the glyphs are placed accordingly, they can either be interpreted individually up close or as a collection from afar. Adding dynamic interaction to these glyphs enhances exploration as users can control the visual representation of information for the desired level of detail. For this reason, we design glyphs to encode the LSRCVT-aggregated information by scales, fields, and spatial regions. 

Overall, we present a novel application of symbolic glyphs which combine scales and fields into a reduced set of visual representations. They are applied within a 3D spatial statistical context to summarize multiscale multivariate information along the surface of a volume. This constitutes a novel multiscale visualization which enables direct comparison of multiple fields and scales with the underlying surface geometry. By linking scale dependence to the local surface topology of flow structures, experts can gain valuable insight on how they behave across scales within a multiphysical space. Our approach extends the LSRCVT-based spatial statistical aggregation method, which did not previously allow for direct visualization of the statistics in a 3D spatial context. This aggregation also enables a data reduction strategy that makes multiscale visualization more feasible across larger datasets.

% Related works section + possible (sub)sections to include
\section{Related Work}
\label{sec:related} 
%R3-C2
In developing our approach, we review the following topological visualizations. \color{black} Heine et al. provide a literature review on state-of-the-art topology-based visualization methods~\cite{https://doi.org/10.1111/cgf.12933}.
Some notable applications include cyclone evolution~\cite{9975824} and fluid flow~\cite{4658165} analysis. The former visualizes cyclone tracks as hierarchical multiscale groups organized by tracking features identified in a merge tree. The latter matches and compares features across two scalar fields based on largest contours, similarity, and locality. Contour trees are then formed and added to the final visualization for ease of navigation. Gyulassy et al. develop a method for topologically robust multiscale analysis~\cite{5228721}. By combining topological concepts with traditional multiscale analysis methods, the key features of structures can be preserved while allowing exploration at varying granularities. 
While we do not explicitly use graph representations, the LSRCVT naturally supports a graph structure formed by a hierarchy of spatial regions~\cite{9904439}. This structure is also compatible with the dynamic nested tracking graph~\cite{8854335}, which tracks and visualizes topological changes over time. Given this, our approach could extend dynamic nested tracking graph-based approaches.\color{black}

We also review the following statistical visualizations. Gosink et al. extend query-driven visualization with a statistical framework~\cite{5473228}, aiming to highlight statistically significant trends within large data. By integrating user-driven segmentation of data, it can display relationships between variables for entries within a select search space. Potter et al. develop EnsembleVis~\cite{5360497} which uses linked interactive overview and statistical displays to visualize ensemble data. The multiple views provide maximal insight while still keeping the information interpretable. Kapelner et al. devise a statistical framework for analyzing cancer cell images~\cite{4272115}. They highlight regions of interest with a segmentation algorithm, displaying the results along with histograms and summary statistics within an interactive visualization system.
We incorporate 2D statistical plots into our system, linking selections there to the corresponding spatial regions in the 3D glyph visualization. With this, domain experts can perform spatially relevant multiscale statistical analysis, aligning with how they conventionally understand these statistics.\color{black} 

Given our approach utilizes spectral scales, we also review applications utilizing the frequency domain in a general sense. Ruzzene devises a frequency-based technique for visualizing damage in structural components, aiming to highlight damage-related signals~\cite{Ruzzene_2007}. The resulting frequency plot can be overlaid onto an image of the structure, showing where and how the damage is affecting it. Matvienko and Krüger develop a frequency-controlled dense flow visualization based on a generalization of the line integral convolution technique~\cite{7429490}. They apply their frequency-filtering method on textured flow data to isolate flow features while providing a clearer view of the original data. In eye-tracking analysis, Li et al. frame the problem of visual saliency in images as a frequency-domain analysis problem~\cite{6243147}. By convolving the image's spectrum with low-pass filters and analyzing the results with a hypercomplex Fourier transform, they are able to predict regions of human fixation across images. Miller et al.~\cite{745302} presents a wavelet transform approach for visualizing text data. The wavelet energy computed from text is analyzed to determine the document's primary thematic characteristics. From there, "islands" are constructed to organize and represent the information. 

\begin{figure*}[!ht] %R2-C1
    \centering
    \includegraphics[width=0.95\linewidth]{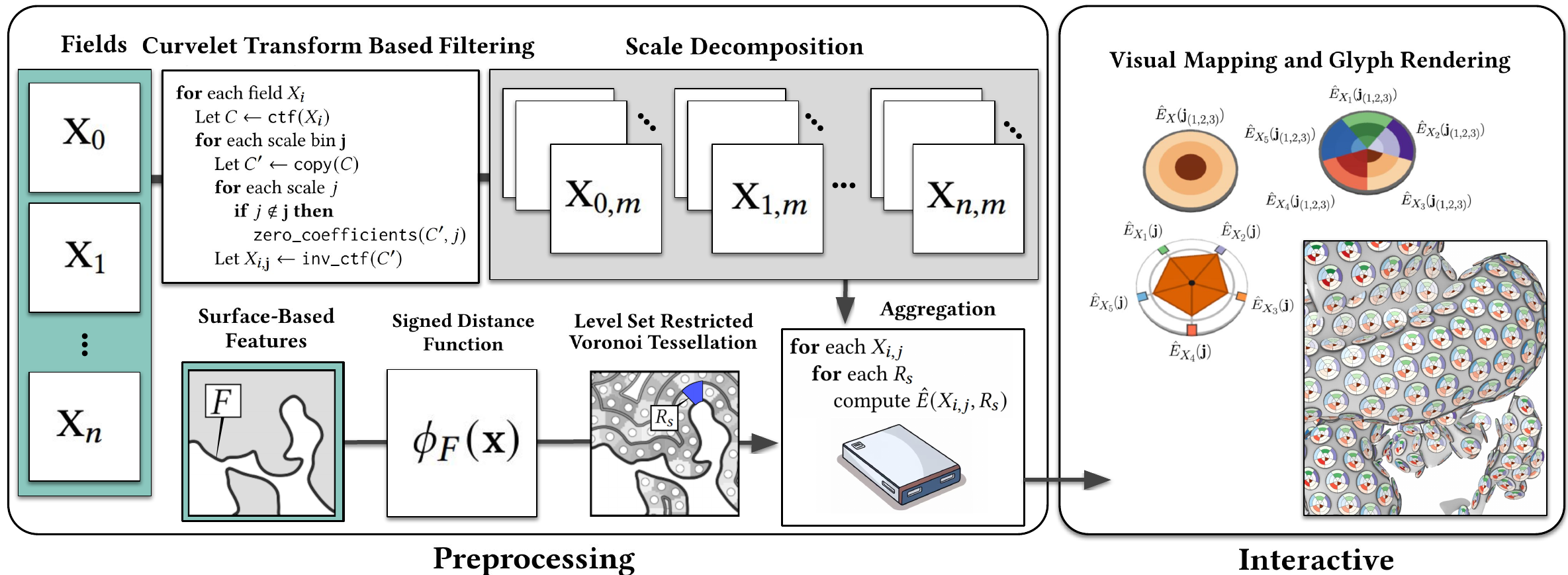}
    \caption{Fields $\mathbf{X}_i$ are input to the scale decomposition, and surface features are input to the LSRCVT. Aggregation of spectral energy is done for each scale of each field based on the LSRCVT regions. Glyphs are generated interactively based on the aggregated data. The top two glyphs show the strength glyph design, the left encoding mean spectral energy for 1 field and 3 scale bins, and the right encoding spectral energy for 5 fields and 3 scale bins. The one below the two shows the starplot glyph encoding 5 fields and 1 scale bin.}
    \label{fig:pipeline}
    \vspace{\figvspace}    
\end{figure*}

Earlier works by Neuroth et al. develop visualization systems based on spatially arranged distribution plots~\cite{7874311,7792155}. Both allow for the navigation of statistical features for ease of analysis. The latter offers multiple plot representations across time and space to highlight prominent features. Dutta et al. also develop a spatiotemporal statistical visualization for anomaly detection in engine blades~\cite{dutta2016situ}. They statistically highlight regions on the blades, corresponding to the parts prone to failing. Bremer et al. introduce a topological approach for hierarchically organizing and navigating large-scale combustion data across time~\cite{bremer2009topological}. Bennett et al. incorporate Bremer's approach to develop a multi-resolution statistical analysis system where feature-based statistics are organized based on spatial and temporal topology~\cite{bennett2011feature}. We build directly off of~\cite{9904439}, which also uses the LSRCVT to spatially aggregate statistics, but does not visualize the aggregated statistic directly in the 3D space or support multiscale analysis. 
Like~\cite{dutta2016situ}, our approach visually maps statistics to their location in space. However, the statistics there are displayed through color-mapped isocontours, which only represent one field, while statistics for multiple fields and scales can be displayed simultaneously through our glyph encoding. Though our visualization does not explicitly display the spatial segmentation of data, like in~\cite{bremer2009topological,bennett2011feature}, the glyphs visuospatially correlate statistics to the centroids of the regions for which they are aggregated, and the segmentation can be configured to provide coarser or finer glyph representations. 
\color{black}

Multiscale visualization and analysis is widely utilized for flow analysis. Cakmak et al. provide a literature analysis on various multiscale visualizations~\cite{9528956}. They highlight the common visualizations, interactions, and applications for visualizing multiscale data, providing a base for developing such visualizations. Bermejo-Moreno and Pullin develop a visualization system to perform multiscale analysis on the geometry of turbulent structures~\cite{BERMEJO-MORENO_PULLIN_2008}. They use the 3D discrete curvelet transform to decompose an incoming scalar field into frequency-component fields, extract structures from each, and then classify them based on the local geometric properties across their surfaces. This forms a geometry-based visualization space where structures can be compared by scale and relative position.
Yang et al. extends the above by applying it to Lagrangian structures~\cite{YANG_PULLIN_BERMEJO-MORENO_2010}. Bussman et al. adapts the same approach to visualize the geometric evolution of turbulent structures~\cite{BUMANN2022105665}.
While these works are focused on geometric analysis, our work focuses more on visualizing multivariate scale dependence and spatial statistical analysis.

Liu et al. develop a multi-objective optimization algorithm for band-pass filter parameters, and use it to apply scale filtering directly in the Kolmogorov energy spectrum. They use their filtering approach to generate visualizations representing one scale band at a time using colored isosurfaces~\cite{https://doi.org/10.1111/cgf.13532}. While our approach also visualizes contributions about a surface, it supports visualizing both multiple fields and multiple scales in one visualization.  

Bermejo-Moreno and Pullin also use the curvelet transform to spectrally decompose flow fields into component fields by length scale. While they utilize the decomposition for geometric analysis of features at different scales, we use multiscale decomposition for analyzing multivariate scale dependence in relation to surface features, like a flame surface\color{black}. 
Additionally, Nguyen et al. propose a novel scale decomposition approach for the visualization of Taylor-Couette turbulence~\cite{9216538}, which is challenging to scale decompose by other filtering methods. Their approach combines multiple attributes into a feature level set and then uses the distance function and kernel density estimation on the feature level set to extract large and small scale structures. They visualize the extracted structures as color encoded isosurfaces, and combine the isosurfaces of multiple scales into a single composite visualization. Our work similarly leverages scale decomposition but instead relies on the CTF. We also visualize multiple scales in one view, but we visualize spectral energy contributions across scale and visualize them with respect to a given surface feature.

For added context, we refer to survey papers on multivariate cross-domain visualization, which our work is one example of.\color{black}\ He et al. and Chen et al. provide surveys on multivariate spatial visualizations~\cite{He_2019,CHEN2019129}. 
He's work identifies feature classification, fusion visualization, and correlation analysis as primary problem domains which help guide the design of multi-physics visualizations.
Meanwhile, Chen's work analyzes visualizations based on the spaces being explored, namely simulation, parameter, and feature space. Dennig et al. build up a framework for dual analysis of feature space and data space based on analyzing existing works~\cite{10158903}.  

%R4-C22
In incorporating multiscale multivariate glyphs, we refer to several works on glyphs to guide our design process. Ward et al. outlines how data can be mapped to graphical attributes in glyphs along with perceptual considerations to take in designing glyphs~\cite{inbook}. Ware also examines glyph design and interaction but with additional emphasis on perception~\cite{WARE2021143}, examining how visual channels are processed hierarchically. These channels can be utilized to highlight prominent features in a visualization, motivating essential design guidelines.
Borgo et al. survey a number of glyph-based visualizations, demonstrating how the aforementioned concepts can be utilized with respect to their application domains~\cite{:10.2312/conf/EG2013/stars/039-063}. %R1-C2,R4-C12
Among these, Ropinski et al. develop an approach for placing and configuring glyphs along isosurfaces such that visual clutter is minimized while maximizing comprehension~\cite{inproceedings}. Ropinski et al. also conduct a survey on glyph-based techniques for spatial multivariate medical data ~\cite{ROPINSKI2011392}. Building off Ropinski's earlier work~\cite{inproceedings}, they evaluate and characterize multivariate glyphs based on preattentive and attentive processing from semiotic theory. We supplement our understanding with Folk and Remington's two models of preattention~\cite{Folk01082006}: a top-down search, where users identify a known pattern, and a bottom-up search, where users identify previously unseen patterns. We utilize the latter conceptualization of preattentive processing. From this, guidelines are developed for integrating glyphs into surface visualizations for visual clarity and coherence. These works guide the design of our glyphs to maximize visual comprehension and visibility with respect to glyphs placed along 3D surfaces. For glyph-based interaction, we also refer to Tong et al.'s GlyphLens which proposes a visually adaptive methodology for glyph-based examination~\cite{7539643}. 
They apply selective distortion upon zoom in order to reveal glyphs occluded by other glyphs. While visual occlusion of glyphs remains a limitation of our work, we could apply similar distortions to mitigate this.

\section{Methodology}
\label{sec:methodology} 

Our visualization pipeline is outlined as follows. First, we perform level-set Voronoi tessellation on a select field to define regions of aggregate statistics along a desired isosurface, and we perform multiscale decomposition for select fields and scales using the curvelet transform. With the data generated from both decompositions, we compute the scale-by-scale spectral energy (i.e. signal contribution) for each field and Voronoi cell. From there, we construct glyphs using the aggregated spectral energy for each cell. Figure~\ref{fig:pipeline} shows an overview of the full pipeline. 

The process begins with a set of raw 3D scalar fields $\mathbf{X}_i$, where $i$ denotes a variable. Each volume $\mathbf{X}_i$ is decomposed by scale using the curvelet transform, generating a new volume $\mathbf{X}_{i,\mathbf{j}}$ for each scale bin $\mathbf{j}$ and field $i$. As discussed in Section~\ref{sec:intro}, we use the curvelet transform (CTF). However, other transforms could be used should they satisfy experts' needs for the data or its application. \color{changed}The main differences and tradeoffs between the CTF and other approaches are discussed by Liu et al.~\cite{https://doi.org/10.1111/cgf.13532}, while Ma et al. provide additional context on the development of the CTF and its advantages over classic spectral methods~\cite{Ma2010TheCT}.\color{black}

Concurrently, we compute the signed distance field for a desired surface which defines the level-set bounded regions to be segmented by the LSRCVT. This segmentation defines the spatial partitioning for all $\mathbf{X}_{i,\mathbf{j}}$ in which local statistics are aggregated. \color{changed} Note that the LSRCVT is a volumetric tessellation. Depending on the distance between level sets, information can be aggregated for the volumetric regions formed in between, rather than directly along, them. Thus, these layers define statistics that are spatially conditioned by the iso-values of level sets. \color{black}While other approaches could be used instead of the LSRCVT, level set restriction is ideal for isolating information near the surface, while the centroidal property ensures uniform glyph placement within that region. A similar algorithm, simple linear clustering~\cite{dutta2017homogeneity}, could be adapted for level set restricted segmentations suitable for aggregation and glyph placement. %A 2D surface-based tessellation could be used as well, though the aggregation would be restricted to a thin sheet of values. 
2D surface-based tessellation is possible, but the aggregation would be restricted to a thin sheet of values.

From this, we have scale decompositions $\mathbf{X}_{i,\mathbf{j}}$ for all fields $i$ and scale bins $\mathbf{j}$, with their statistical aggregations. Collectively, $\mathbf{X}_{i,\mathbf{j}}$ incurs high storage costs but the aggregations, from which visualizations are generated, are comparatively lightweight. Thus, each $\mathbf{X}_{i,\mathbf{j}}$ could be discarded after aggregation to ease those costs. However, additional aggregations cannot be computed thereafter, as the original data is required. Thus, it is wise to generate multiple aggregations across different tessellations beforehand to support multi-resolution visualization with lower storage costs. Likewise, the preprocessing stages execute relatively slowly. Once completed, however, the data can be efficiently and effectively visualized with our interactive system.

\begin{figure}[h]
    \centering
    \includegraphics[width=1.0\linewidth]{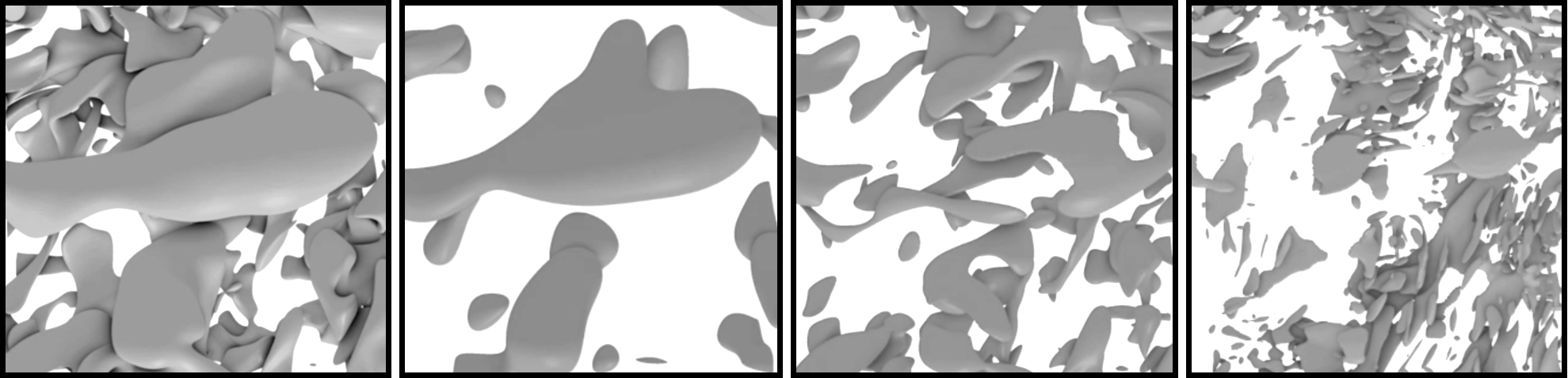}
    \caption{Isosurfaces of a scale decomposition: (left to right) input field, large, medium, then small scales.}
    \label{fig:scales}
    \vspace{\figvspace}     
\end{figure}

\subsection{Curvelet Transform Multiscale 
Decomposition}\label{ctfmd}
The curvelet and curvelet transform (CTF) were conceived by Emmanuel Candes and David Donoho as a means of detecting curved edge discontinuities ~\cite{doi:10.1137/05064182X,https://doi.org/10.1002/cpa.10116}. Based on this, curvelets are defined as signals with %frequency 
scale component $j\in \mathbb{N}$, orientation component $l\in \mathbb{N}$, and spatial translation component $\mathbf{k} \in \mathbb{N}^3$. In the spatial domain, these curvelets manifest as needles or ridges with a size and tilt based on their $j$ and $l$ values respectively and are positioned based on their $\mathbf{k}$ coordinates. For multiscale analysis, we focus on the curvelet scales $j$. In the discrete curvelet transform, each scale $j=j_0, j_1,...,j_{e}$ isolates features based on a Cartesian tiling of the Fourier space. There, they correspond to concentric square windows which isolate progressively finer scales for increasing $j$. In turn, the spatial length of the curvelets decrease by a factor of 2 per $j$. Because of this, the maximum number of scales is $log_2(N/2)$, where $N$ is dimension of the incoming volume in grid units. The physical length of the input is $L=N\Delta x$, where $\Delta x$ is the grid unit length. This allows dimensionless curvelet lengths, expressed as a unit fraction, to be converted to physical ones. E.g., the dimensionless length $1/2^j$ becomes $L/2^j$.

%R4-C4
Thus, the curvelet transform takes a scalar field and decomposes it into its component curvelets and their coefficients, denoting signal contribution. We utilize Lexing Ying's \textsc{fdct3d} implementation of the 3D fast discrete curvelet transform for the multiscale decomposition~\cite{candes2005curvelab}\color{changed}. \color{black}
%~\footnote{https://curvelet.org/software.php}. 
To do this, we first pass the scalar field through the forward transform to get its curvelet coefficients for each scale. Then, we selectively filter out scales by zeroing out coefficients in the irrelevant scale bins. This is to extract and isolate flow features at particular length scales. In the case of combustion analysis, for example, we might only want to look at features in flame length scale. \color{black} Finally, we pass the filtered coefficients through the inverse curvelet transform to obtain the component fields, i.e. the flow structures corresponding to the selected scales. Figure~\ref{fig:scales} shows the resultant isosurfaces from this decomposition.\color{black} 

\begin{figure}[h]
    \centering
    \includegraphics[width=0.95\linewidth]{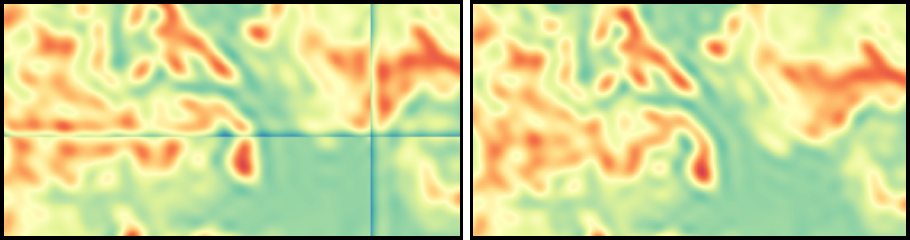}
    \caption{(Left) Border artifacts introduced after spectral filtering. (Right) Artifacts resolved by using overlapping windowed blocks.}
    \label{fig:artifacts}
    \vspace{\figvspace}     
\end{figure}

\subsubsection{CTF Preprocessing} %R4-C9
\label{ctfpp}

As with standard DFT-based spectral processing, the 3D discrete CTF implementation we use, which implicitly utilizes DFTs, assumes that the incoming scalar field has periodic boundary conditions. Yet, many lack this property. Thus, additional preprocessing may be needed to avoid spectral leakage and boundary artifacts within the CTF output. One way is through mirror extending the data, which pads each side with values reflected along the boundary. This mimics periodicity along all dimensions, which can mitigate the propagation of sharp boundary artifacts in the volume. However, this increases the size of the volume to be decomposed by a factor of 8. Application of a cosine transform-based CTF was considered as a less storage-intensive alternative, since it is functionally equivalent to applying the CTF on mirror-extended data. An implementation of such is even outlined by Demanet and Ying~\cite{Demanet2007CurveletsAW}. However, it can only decompose 2D input. A more viable approach applies windowing functions to smoothly roll off values along the boundaries, ensuring continuous wrapping there. Thomson et al. use this for parallel block-based CTFs where a cosine taper (i.e. Tukey window) is applied over a fixed region around the edges of overlapping blocks. These blocks are then composited together such that the taper is removed, creating a virtually seamless result~\cite{thomson2006parallel}. Figure~\ref{fig:artifacts} compares the result of this CTF with a non-windowed block CTF. Mirror extension is used on smaller chunks of data ($256^3-512^3$) while the parallel windowed block approach is used on larger datasets.

From this, there are some issues to account for. One is any border discontinuities introduced from clipping or padding. Another is the removal of high-frequency information from downsampling, as the fidelity of the grid determines the smallest scales that can be resolved, which is 2 times the length of grid spaces. In computing spectral energy for each scale of our combustion data, for example, we found that the smallest curvelet scale contained negligible spectral information. This is because the base grid resolution is higher than the size of its finest structures for simulation accuracy. Thus, we were able to downsample by a factor of 2 for efficient data processing, with little loss of information. In any case, we consulted with relevant experts on how to best preprocess their datasets for spatially and physically accurate CTF decompositions. 

After processing the fields to satisfy input requirements, the CTF multiscale decomposition can be configured \color{changed} and \color{black} then applied to spectrally filter them by scale bins as described in Section 3.1, where bins are precomputed based on the experts' desired ranges of frequency or length scales for visualization. This yields additional data from which multiscale statistics are aggregated in Section 3.3.\color{black}
\subsection{Level Set Restricted Voronoi Tessellation}
\label{sec:tessellation}
\begin{figure}[h]
    \centering
    \includegraphics[width=0.95\linewidth]{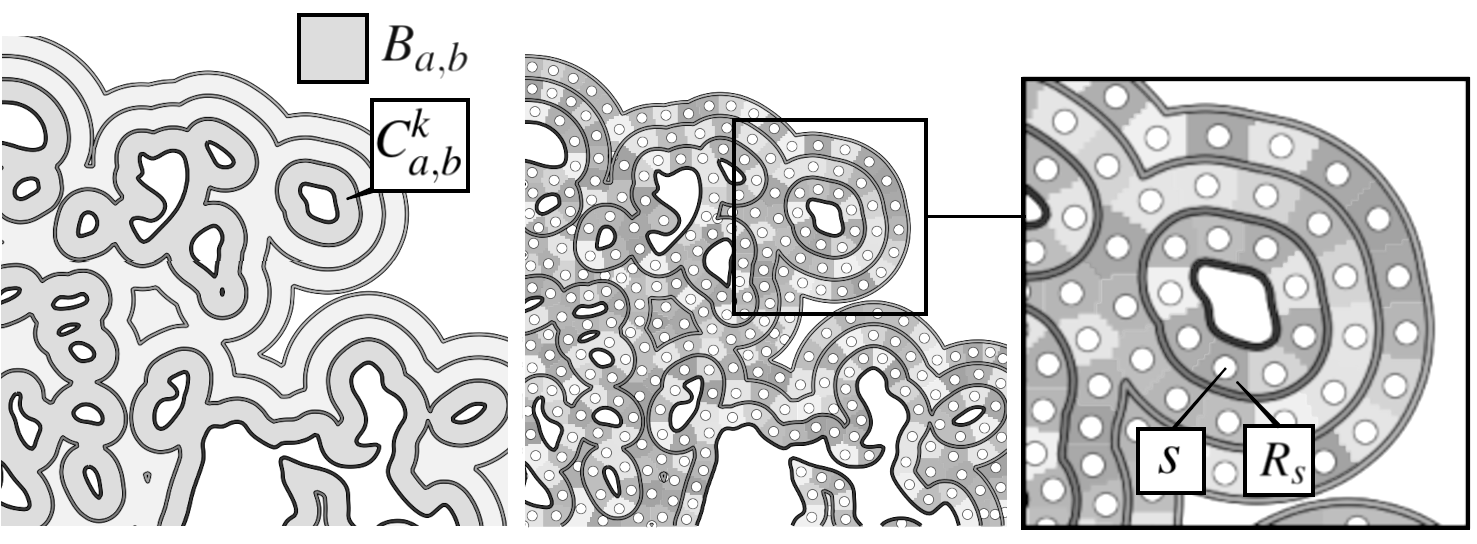}
    \caption{A 2D LSRCVT using the distance field. (Left) The darker region highlights band $B_{a,b}$. $C_{a,b}^k$ denotes a component of $B_{a,b}$. (Right) Elements in $R_s$, where $R_s \subset C_{a,b}^k \subset B_{a,b}$, are aggregated. %From there, 
    The aggregate data can be encoded as a glyph centered at site $s$.}
    \label{fig:cellular}
    \vspace{\figvspace}    
\end{figure}

The LSRCVT decomposes a spatial domain into a hierarchical set of regions~\cite{9904439}. The first level is specified based on a sequence of isocontours with isovalues, $[c_0, c_1, ..., c_n]$, which define volumetric bands $B_{c_i,c_{i+1}}$. The bands are then broken down into connected components $C_{a,b}^k$ where $B_{a,b} = C_{a,b}^0 \cup C_{a,b}^1 \cup ... \cup C_{a,b}^m$, and no element of $C_{a,b}^i$ is spatially adjacent to any element in $C_{a,b}^j$ for any $(i,j)$. Then, the connected components are tessellated with a centroidal Voronoi decomposition. Each centroidal Voronoi region is situated about a site point $s=(x,y,z)$. Such region $R_s$ contains all elements which are both within the same connected component as $s$ and are closer to $s$ than any other Voronoi site $s_j$ in the same component. The decomposition is then used to aggregate data within each region. Figure~\ref{fig:cellular} illustrates this with a simple example in 2D. 

%R4-C19; R4-C8, C18
For the LSRCVT-based glyph visualization, we first select a surface-based feature $F$, then compute the signed distance field $\phi_F(\mathbf{x})$. Next, we choose a set of distance values to form the bands (i.e. layers surrounding $F$) used for the LSRCVT. In addition to these level-set values, we select a value for Voronoi site density, which also determines glyph density. Sparser glyph sets enable larger glyph sizes, but with coarser spatial aggregation. Also, if glyphs are too large, they may not align with the surface, even intersecting it, should it have small noisy features. To enhance glyph legibility, multiple tessellations can be computed with increasing glyph density, corresponding to increasing spatial resolution for statistical aggregation and glyph representation.

%R4-C20
To compute $\phi_F(\mathbf{x})$, we use Fernandez's \textsc{TriangleMeshDistance} Library~\cite{JAFICG}
% ~\footnote{https://github.com/InteractiveComputerGraphics/TriangleMeshDistance},
and to compute the LSRCVT, we use an 
%open source 
implementation in the \textsc{Marrus} project~\cite{marrus}.
%~\footnote{https://gitlab.com/tneuroth/marrus}. 
For larger data, we use the multiblock version of the LSRCVT which computes the tessellation in parallel over blocks and then merges the result.\color{black}
\subsection{Multiscale Aggregation}
\label{sec:aggregation}
After multiscale decomposition, we aggregate over Voronoi regions. Since we may combine multiple curvelet scales in one bin, we denote $\mathbf{j} = [ j_{\min}, j_{\min}+1, ..., j_{\max}]$, with $j_{\min}>j_0$ and $j_{\max} \leq j_e$ representing the edges of the merged curvelet scales. For each bin, we compute mean spectral energy $\hat{E}_X(\mathbf{j})$ per region $R_s$. Since the bins span unequal scale ranges, we also divide by a factor, $\Delta \mathbf{j} = \frac{L}{2^{j_{\min}}} - \frac{L}{2^{j_{\max}+1}}$, where $L$ is the domain length, which approximates average contribution per unit length scale. The formula is
\vspace{-0.15em}
\[\hat{E}_X(\mathbf{j}) =  \frac{1}{ n \Delta \mathbf{j} }\sum_{v \in R_s} |x_\mathbf{j}(v)|^2,\] 

where $n$ is the number of voxels in $R_s$. Note that $j_0$ corresponds to a low-pass window with an undefined maximum length scale. As a convention, one can use $L$, consistent with the above formula. However, the interpretation of bins including $j_0$ will depend on the length of the grid. To ensure stable results across different input sizes, one may simply exclude $j_0$.

%R4-C5
Depending on the use case, the norm, $|x_\mathbf{j}(v)|$, may be used instead of $|x_\mathbf{j}(v)|^2$. The latter gives the spectral energy per its definition in signal processing, in line with our combustion experts' understanding of multiscale turbulence behavior. In contrast, our channel flow expert preferred using the unsquared norm to compare the spectral contribution of terms within a differential equation. \color{black} Select scales can also be converted to other analytically relevant units (e.g. characteristic flame length scale for combustion analysis).

%R4-C4,C10
The aggregation of scales into bins is primarily motivated by the combustion scientists to hone analysis on the most relevant ranges of scales. They determined bins based on lengths proportional to the characteristic flame scale for their simulation. For channel flow analysis, a larger number of bins were constructed for a more scale-refined analysis, per that expert's needs. In general, fewer scale bins make the glyphs more legible but provide less insight on how spectral contribution balances across each scale. \color{black}

\subsection{Multivariate Glyphs}
\label{sec:glyphs}
%With the data obtained in 
Following the previous steps, we place glyphs at Voronoi cell sites. Two glyph designs are developed, offering multiple representations of multiscale field data for comparative analysis across regions. 

\begin{figure*}[!ht] %R4-C16
    \centering
    \includegraphics[width=0.9\linewidth]{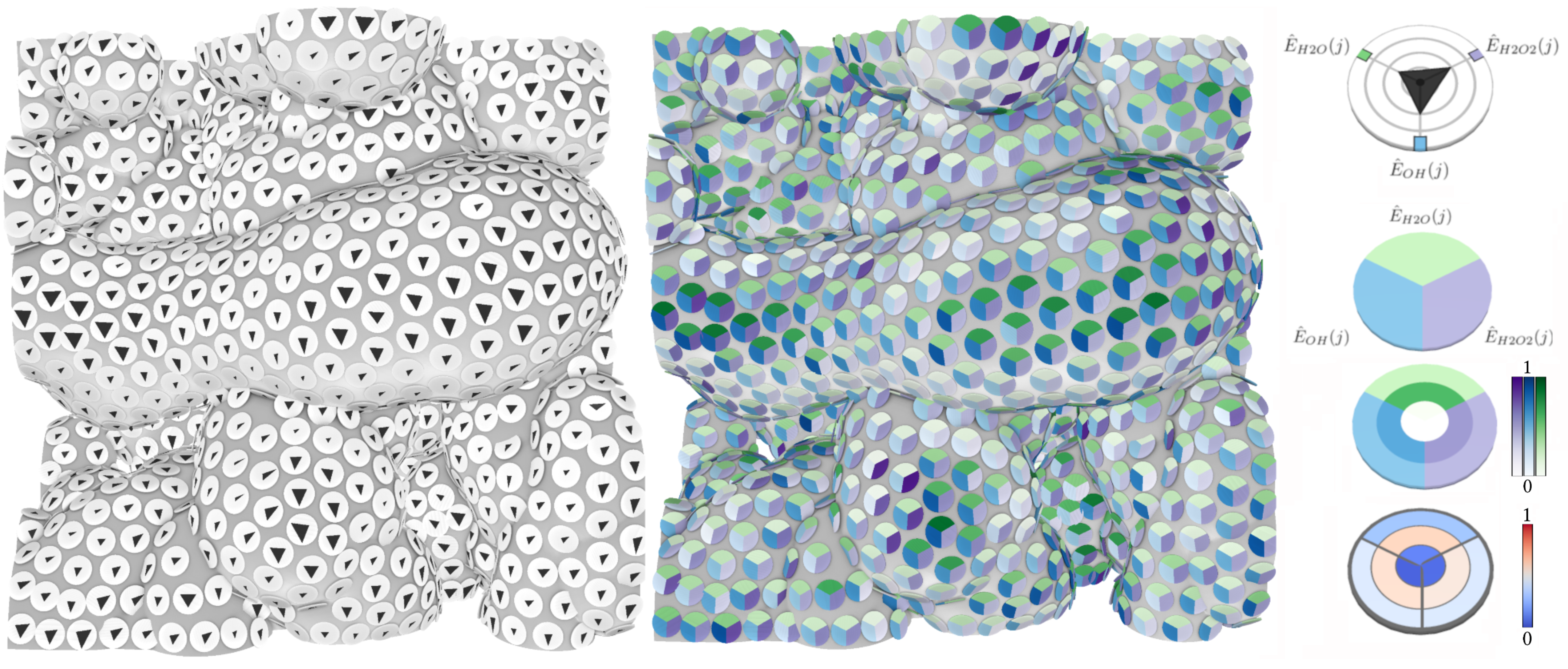}
    \caption{The two %views on the left 
    visualizations 
    show the same information, spectral energy contribution of the highest frequencies for 3 variables, using the starplot glyph (left) and strength glyph (right). The user can interactively switch between the two glyph types. The top two glyphs on the right show averages over all glyphs. While color-based encoding makes it easy to see variation from afar, a user's perceptual mapping between colors to values can be more complex or error-prone since it requires referencing color legends, and depends on the user's color perception. Distance-based value encoding tends to be perceived more intuitively and precisely. The bottom two glyphs %on the right 
    show mean spectral energy for 3 variables and 3 scale bins. Using separate color maps for each variable makes it easier to tell different variables apart at a glance, but makes comparing values across variables more difficult. The bottom %right 
    glyph uses the same color map for all 3 variables to make comparison easier.}
    \label{fig:compare_glyphs}
    \vspace{\figvspace}    
\end{figure*}

\paragraph*{Strength glyph.} The strength glyph is inspired by the strength disks seen in Carlos Costa's "Demystifying Curvelets"~\cite{towardsdatascienceDesmystifyingCurvelets}. As shown in Figure~\ref{fig:pipeline}, the primary characteristic of this glyph is its concentric rings, where each ring encodes the contribution for a scale or range of scales. The rings closer to the center denote coarser scales while rings further from the center denote finer scales. Each ring is colored based on the $\hat{E}_X(\textbf{j})$ computed for field $X$ and scale bin $\textbf{j}$ of the enclosing region. To encode scale-by-scale contribution across multiple fields, these rings can be divided into sectors as shown in the right of Figure~\ref{fig:pipeline}, where their color encodes field $X_i$ while the rings encode $\hat{E}_{X_i}(\textbf{j})$. Each wedge in the strength glyph is assigned its own color map where $\hat{E}_{X_i}(\textbf{j})$ is normalized between 0 and 1. Users can select the color map for each wedge. This glyph is designed to take advantage of concentricity, shape, and color to resist orientation-based deformation while viewing. Because the encoded scale bins are arranged as concentric rings atop a circular disk, the placement of those bins is consistent from almost any viewing angle, except for when the disk is laid flat towards the viewer. Thus, the larger-scale representations will always be placed towards the center of the disk while smaller-scale representations will always be placed further out from the center. No matter what color map is used, the color assigned to each ring enforces consistent viewing by preserving the locality of scale bins defined by the rings. The same does not always apply to wedges depending on what color map is used. If different color maps are used for each wedge, then it is clear which wedge corresponds to which field. In the case where the same color map is used across all wedges, then it becomes more important to keep the orientation of the glyph consistent with respect to the viewer. In either case, the ring segments and their colors can form unique and identifiable shapes and patterns based on their arrangement and contrast. The main drawback of this representation comes from the number of small visual subelements introduced as more wedges are added. This multi-wedge multi-ring glyph displays a lot of information up close, making it good for attentive processing but becomes hard to discern and differentiate when these glyphs are collectively viewed from afar, making it less ideal for preattentive analysis. To counter this, users can change the number of wedges and rings shown, allowing them to adjust the visual complexity of glyphs for preattentive or attentive analysis. Simpler configurations of this glyph aim to facilitate the "pure-capture" model of preattention, where glyphs are targeted for examination based on the most prominent patterns formed. The more detailed configurations are less suitable as they introduce too much visual noise. This is in contrast to the "contingent-capture" model, where users would search for one or more target glyphs. While this kind of analysis could be done for simpler glyph configurations, we design all glyphs with the former model in mind, as it 
better reflects expert usage.
%more closely aligns with how experts would use them.
\color{black}%(R4-C12)

\paragraph*{Starplot glyph.} 
The starplot glyph consists of a starplot polygon laid atop a disk. The polygon itself has 3 or more points representing $\hat{E}_{X_i}(\textbf{j})$ for each field $X_i$ at fixed scale bin $\textbf{j}$ (see Figure~\ref{fig:compare_glyphs}). We considered having multiple starplot polygons representing  $\hat{E}_{X_i}(\textbf{j}')$ for other scale bins $\textbf{j}'$ allowing for multiscale analysis within a single representation. However, we found that the overlap would obfuscate the polygons, making it harder to identify distinct shapes from afar. Users can instead change the polygon shown on the disk based on which scale bin they wish to analyze. Each scale bin would be assigned different polygon colors. This glyph aims to take advantage of the Gestalt principles of symmetry and Prägnanz, i.e. viewer's preference towards symmetry and simplicity in visual representations. Since starplots can form a variety of shapes with differing amounts of symmetry or convexity, viewers can identify a collection of polygons that look more or less symmetric or polygons that are more convex or concave. While the starplot can support preattentive processing this way, the polygon is subject to visual deformation if not oriented towards the viewer. We also lose the ability to visualize multiple scales in the same glyph from afar. While multiple fields could be visualized effectively in the starplot glyph from close up, to enable visualization of each scale bin and field at the same time from afar, we leverage multiple linked views. For each view, the starplot glyphs can be configured to show contributions at different scales. \color{black}%(R4-C14)

%\paragraph{\textbf{Vector glyph}.} The vector glyph consists of three dimensional vectors with color, length, and direction. As shown in Figure~\ref{fig:vector_glyph}, each colored arrow corresponds to a select scale bin. In particular, each vector represents the vector sum of mean spectral energy across three directional fields $v_x$, $v_y$, and $v_z$. The length of these vectors are normalized between 0 and 1 such that each scale vector can be compared with respect to their relative contribution. This glyph takes advantage of the characteristics seen in traditional vector or arrow glyphs (e.g. hedgehog glyphs, vector fields), namely the arrow's length, the arrowhead's position, and the color. In having the three arrows joined as a singular representation, the average direction of contribution across scales can easily be compared within a single visual representation. Likewise, the color allows each arrow to be distinguished, regardless of viewing angle or orientation. Similarly to the strength glyph, however, this glyph does not fare as well when viewed from afar as the conjoined arrows are harder to distinguish. To counter this, users can add or remove arrows from the glyph, allowing them to adjust the visual complexity of glyphs for preattentive or attentive analysis.

\subsection{Rendering}
\label{sec:rendering}

The glyphs are placed homogeneously around surface-based flow features, one layer at a time. Rendering the surface beneath the glyphs adds geometric context while reducing visual clutter from overlapping glyphs. Ambient occlusion is used to obtain an auxiliary shadow effect that helps with depth perception.

The rendering and visual encoding of the glyphs is done through a 3D graphics pipeline using instanced rendering of a base glyph model, each instance of which is transformed and visually encoded inside the vertex and fragment shader functions.

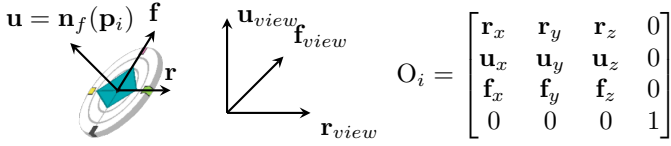
\begin{figure}[h]
\begin{tabular}{@{}m{0.6\columnwidth}@{}m{0.3\columnwidth}@{}}
\begin{tikzpicture}[scale=1.5]
    \draw[black, thick, -stealth] (0,-0.3) -- (0.74,-0.3) 
        node[below right] {$\mathbf{r}_{view}$};
    \draw[black, thick, -stealth] (0,-0.3) -- (0.5,0.2) 
        node[above right] {$\mathbf{f}_{view}$};
    \draw[black, thick, -stealth] (0,-0.3) -- (0,0.54) 
        node[right] {$\mathbf{u}_{view}$};
    \node[inner sep=0pt] at (-0.95,-0.13) {\includegraphics[width=0.075\textwidth]{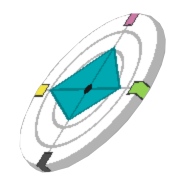}};
    \draw[black, thick, -stealth] (-0.97, -0.1) -- (-1.39, 0.32 ) 
        node[above] {$\mathbf{u}=\mathbf{n}_{f}(\mathbf{p}_i)$};
    \draw[black, thick, -stealth] (-0.97, -0.1) -- ( -0.5, -0.1 ) 
        node[above] {$\mathbf{r}$};
    \draw[black, thick, -stealth] (-0.97, -0.1) -- ( -0.64, 0.43 ) 
        node[above] {$\mathbf{f}$};
\end{tikzpicture}
& 
 \begin{minipage}{0.4\textwidth}
            $\displaystyle
            \mathbf{\mathrm{O}}_i = 
            \begin{bmatrix}
            \mathbf{r}_x & \mathbf{r}_y & \mathbf{r}_z & 0 \\
            \mathbf{u}_x & \mathbf{u}_y & \mathbf{u}_z & 0 \\
            \mathbf{f}_x & \mathbf{f}_y & \mathbf{f}_z & 0 \\
            0 & 0 & 0 & 1
            \end{bmatrix}
            $\\ \\ \\ \\
        \end{minipage}
\end{tabular}
\vspace{-0.4in}
    \caption{An illustration of the change of basis transformation that reorients each glyph so that its right vector $\mathbf{r}$ is coplanar with the $\{\mathbf{r}_{view}, \mathbf{f}_{view}\}$ plane.}
    \label{fig:transform}
    \vspace{-0.1in}
\end{figure}

\begin{figure*}[t]
      \centering
      \includegraphics[width=0.93\linewidth]{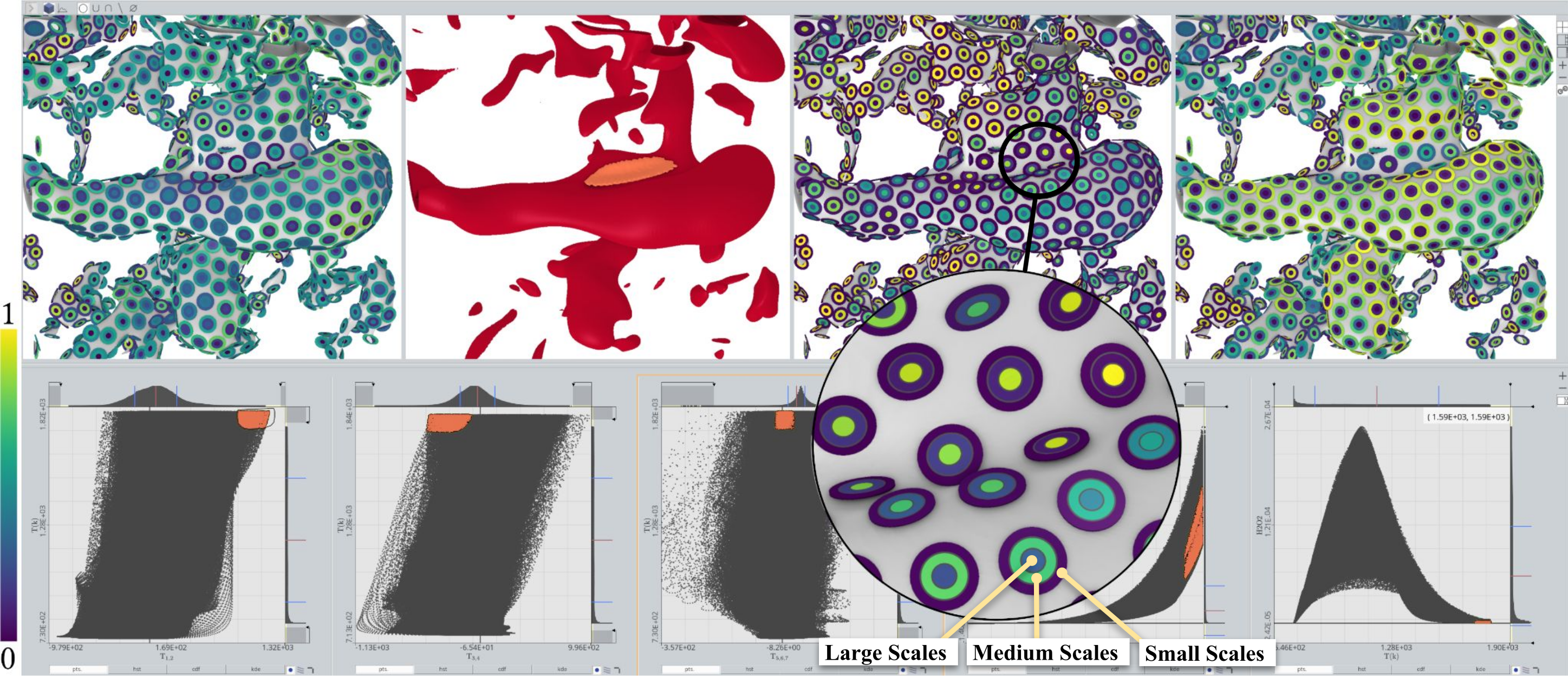}
      \caption{Top row from left to right: OH mass fraction, temperature isosurface, temperature, H$_\mathrm{2}$O$_\mathrm{2}$ mass fraction with glyphs representing contributions from scale bands. Bottom row from left to right: temperature vs. large scale, intermediate scale (scale close to that of an unperturbed flame) and small scale temperature, OH mass fraction vs. temperature and H$_\mathrm{2}$O$_\mathrm{2}$ mass fraction vs. temperature. \color{changed} The zoomed-in region shows glyphs laid across a concave flame element (highlighted in orange on the temperature isosurface). Note the difference in appearance between glyphs in the cavity vs those outside.\color{black}}
      \label{fig:case_study_1_2}
      \vspace{\figvspace}    
\end{figure*}

To maintain visual perception of the surface features, we rotate the glyphs according to the normal vectors along the surfaces of interest. We also rotate the glyphs so as to maintain a consistent orientation relative to the viewer. More precisely, the glyph's up vector is the surface normal at the position of the glyph, $\mathbf{n}_f{(\mathbf{p}_i)}$, while its right vector is made parallel with the $\{\mathbf{r}_{view},\mathbf{f}_{view}\}$ view plane, where $\mathbf{r}_{view}$ and $\mathbf{f}_{view}$ are the view's right and forward vectors respectively. These vectors are depicted in Figure~\ref{fig:transform}. The transformation happens in the vertex shader using a look-at matrix $\mathrm{\mathbf{O}}_i$, which performs a change of orthogonal basis. This matrix is constructed for each glyph instance $i$ as follows. Given the surface normal $\mathbf{n}_f$ at glyph instance position $\mathbf{p}_i$ and view matrix $\mathrm{\mathbf{V}}$, the up vector of the glyph is defined as $\mathbf{u} = \mathbf{n}_f{(\mathbf{p}_i)},$ while the up vector of the view is defined as $\mathbf{u}_{view} = \mathrm{\mathbf{V}}^{\top} \mathbf{(0,1,0)}$. From there, the right vector of the glyph is computed as $\mathbf{r} = \mathbf{u} \times \mathbf{u}_{view}$ from which the forward vector of the glyph is computed as $\mathbf{f} = \mathbf{r} \times \mathbf{u}$. Finally, the orientation matrix for glyph instance $i$ is shown in Figure~\ref{fig:transform}.

The data that is encoded in the glyphs is accessed in the shader functions using a texture buffer, \vspace{-0.15em} \[
\mathbf{T} = \left[ s_0^i(f_0), s_1^i(f_0), \dots, s_n^i(f_0), s_0^i(f_1), \dots, s_n^i(f_m) \right],
\] where $s_j^i(f_l)$ is the $k^{th}$ scale bin of the $l^{th}$ field within the $i^{th}$ Voronoi region. We can then access the data in the shader based on the built-in variable that gives the vertex instance corresponding to the Voronoi site index. In OpenGL, the variable is called \verb|gl_InstanceID|. This variable is also passed along to the fragment shader which will use the glyph data to do the visual encoding. To access $s_j(f_l)$ for glyph $i$, the following formula is used: $ s_j^i(f_l) = \text{texelFetch}( \mathbf{T}, i \cdot M \cdot N + N \cdot l + k )$.

The visual encoding depends on the selected glyph design but is generally done in the fragment shader on a per-fragment basis. There, the position and orientation of the fragment relative to the base glyph model are used to determine if it is associated with the background, glyph border, or glyph interior. If the fragment is associated with a glyph interior, then the field and bin that the fragment corresponds to are determined. First, the position $\mathbf{p}_m$ in model space is passed from the vertex shader to the fragment shader. Since the base instance model is centered in model space at $(0,0,0)$ and scaled to size $(1,1,1)$, we can get the angle about $\textbf{u}$, $\theta = \text{atan}( \mathbf{p}_m.x, \mathbf{p}_m.z ) + \pi$, and radius $r=\text{length}(\mathbf{p}_m)$.

For the strength glyph, wedges correspond to a field and are partitioned into concentric rings corresponding to the scale bins of the field. The color of the ring is used to encode the scale bin's contribution. Thus we obtain the bin index $k$ and field index $l$ of the fragment using their formulae $k = \min(N - 1, \lfloor r \cdot N \rfloor)$ and $l=\min(M - 1, \lfloor \theta \cdot M / (2\pi) \rfloor)$. From here, we access the value of $s_j(f_l)$ in the texture buffer using its aforementioned formula. Optionally, square root or logarithmic scaling can then be applied, which can help to perceive variation in the smaller values that might otherwise be suppressed in the encoding (e.g. wide range of magnitudes, extreme outliers).%R4-C21

% Clamping to a custom range would be another approach to handling unwanted effects introduced by extreme outliers.

\color{black}

In the starplot glyph, the glyph is divided into radial axes for each field. The distance from the origin along the axis encodes the contribution of the bin for its corresponding field. Connecting these points forms wedges between the field axis $l_a$ and its neighboring field axis $l_b = (l+1) \mod M$. Together, these wedges compose a polygon. To render the wedges, we first determine which of these wedges the fragment belongs to, using the formula for $l$ to get $l_a = \min(M - 1, \lfloor \theta \cdot M / (2\pi) \rfloor)$. To determine if the fragment is within the wedge of the polygon, we first construct points $\mathbf{p}_a = ( \text{cos}( \theta_a ), \text{sin}( \theta_a ) )$ and $\mathbf{p}_b = ( \text{cos}( \theta_b ), \text{sin}( \theta_b ) )$, where $\theta_a = l_a \cdot 2 \cdot \pi / M $ and $\theta_b = l_b \cdot 2 \cdot \pi / M$ are the angles of the axis for the two fields about the origin. Then, we do a simplified point-in-triangle test to see if the fragment position $\mathbf{p}_m$ lies within the triangle formed by the coordinates $\left[ \mathbf{p}_a, \mathbf{p}_b, (0,0) \right]$. If $$(\mathbf{p}_b.x - \mathbf{p}_a.x) (\mathbf{p}_m.y - \mathbf{p}_a.y) - (\mathbf{p}_b.y - \mathbf{p}_a.y)  (\mathbf{p}_m.x - \mathbf{p}_a.x) > 0,$$ then the fragment is in the wedge. We can then output the fragment color accordingly. Other tests to determine if the fragment belongs to a border, visual axis, grid line, or other visual annotations on the glyph follow similarly to what is described above. Thus, we omit the details of those tests for brevity.

To optimize the perceptible dynamic range in the visual encodings while comparing across fields and spatial regions with widely varying value magnitudes, having different normalization options is beneficial. One option is \textit{(global $\mid$ local)} spatial normalization, or \textit{(GSN $\mid$ LSN)} for short. \textit{GSN} normalizes based on the minimum and maximum values over all of the spatial bands in the tessellation while \textit{LSN} normalizes only with respect to the band displayed in the single view. The other option is \textit{(global $\mid$ local) scale} bin normalization, or \textit{(GBN $\mid$ LBN)} for short. \textit{GBN} normalizes based on the minimum and maximum values over all scale bins per field while \textit{LSN} normalizes based only on the currently displayed scale bins. When comparisons are done across views and spatial regions or scale bins, it is useful to use the global options so the visual encodings in each view are identically calibrated. We denote the scalar mapping range for the visual encoding of field $f$ as 
\[
\gamma_l = \left( \min_{\mathbf{j}, \mathbf{R}_i} \, s_\mathbf{j}^i(f_l), \, \max_{\mathbf{j}, \mathbf{R}_i} \, s_\mathbf{j}^i(f_l) \right), \quad \text{for } (\mathbf{j}, \mathbf{R}_i) \in \mathbf{J}' \times \mathbf{B}'.
\]
where \(\mathbf{R}_i\) are the Voronoi regions, \(\mathbf{j}\) are the scale bands, and
\begin{align}
\label{eq:1}
\mathbf{B}' = 
\begin{cases}
    \mathbf{B}, & \text{if GSN} \\
    \mathbf{B}_{\text{vis}} \subseteq \mathbf{B}, & \text{if LSN}
\end{cases} \hspace{1em}
\mathbf{J}' = 
\begin{cases}
    \mathbf{J}, & \text{if GBN} \\
    \mathbf{J}_{\text{vis}} \subseteq \mathbf{J}, & \text{if LBN}.
\end{cases}
\end{align}
\vspace{-0.175em}
where $\mathbf{B}_{\text{vis}}$ is the subset of spatial bands and $\mathbf{J}_{\text{vis}}$ is the subset of scale bins that are currently selected for the visualization in view.

The third \textit{per-glyph normalization} (\textit{PGN}) option normalizes bins per-glyph and per-variable so that their values sum to 1 per region. Likewise, we set $\hat{s}^i_{\mathbf{j}} (f_l) = \left(s^i_{ \mathbf{j} } (f_l) \right) / \left( \sum_{R_i} s^i_{\mathbf{ j } }(f_l)\right)$. With this option, the overall balance of scale contributions can be more easily perceived within each individual glyph. The fourth \textit{all axes} option normalizes each axis identically. Finally, the fifth option fixes the minimum of the scalar mapping range to 0.

\subsection{Integrating Data into Visualization} 
\label{sec:vis-system}

The glyphs described in the previous sections are integrated into the system as follows. A configuration is loaded specifying the glyph data, parameters, and encoding, along with information for additional visualizations, such as isosurfaces and 2D statistical plots. Once loaded, users have access to an interface consisting of a 3D rendering panel, a statistical plot panel, and a side panel for configuring the views. The statistical plot panel provides a statistical overview of the incoming data with 2D scatterplots along with the 1D histograms for each axis. These plots enable users to select and highlight specific subsets of the data, displaying the given selection within the statistical and spatial views. Meanwhile, the 3D rendering panel displays level-set surfaces with their corresponding glyphs along with the isosurfaces specified in the configuration file, providing the spatial representations the incoming data. From there, additional 2D plots and 3D views can be added to the corresponding panels and then configured separately using the side panel. 

There, users can configure the type, size, and appearance of glyphs as described in Section~\ref{sec:glyphs} and ~\ref{sec:rendering}. Users can also toggle the visibility of loaded tessellation bands and isosurfaces. In addition, users can enable the display of select statistics within the 3D view, corresponding to their spatial location. This selection is made using a lasso tool on a chosen 2D plot and is linked to a select 3D view. The selection is then highlighted in the linked 3D view and across the corresponding regions of all other statistical plots. With this, users can further correlate and contextualize the local statistics with spatial representations of the same information. Figure~\ref{fig:case_study_1_2} shows a snapshot of the system, the implementation of which is open source as part of \textsc{Marrus}~\cite{marrus}. Detailed system interactions are shown in the accompanying video~\cite{demo-video}.

\begin{figure*}[t]
      \centering
      \includegraphics[width=1.0\linewidth]{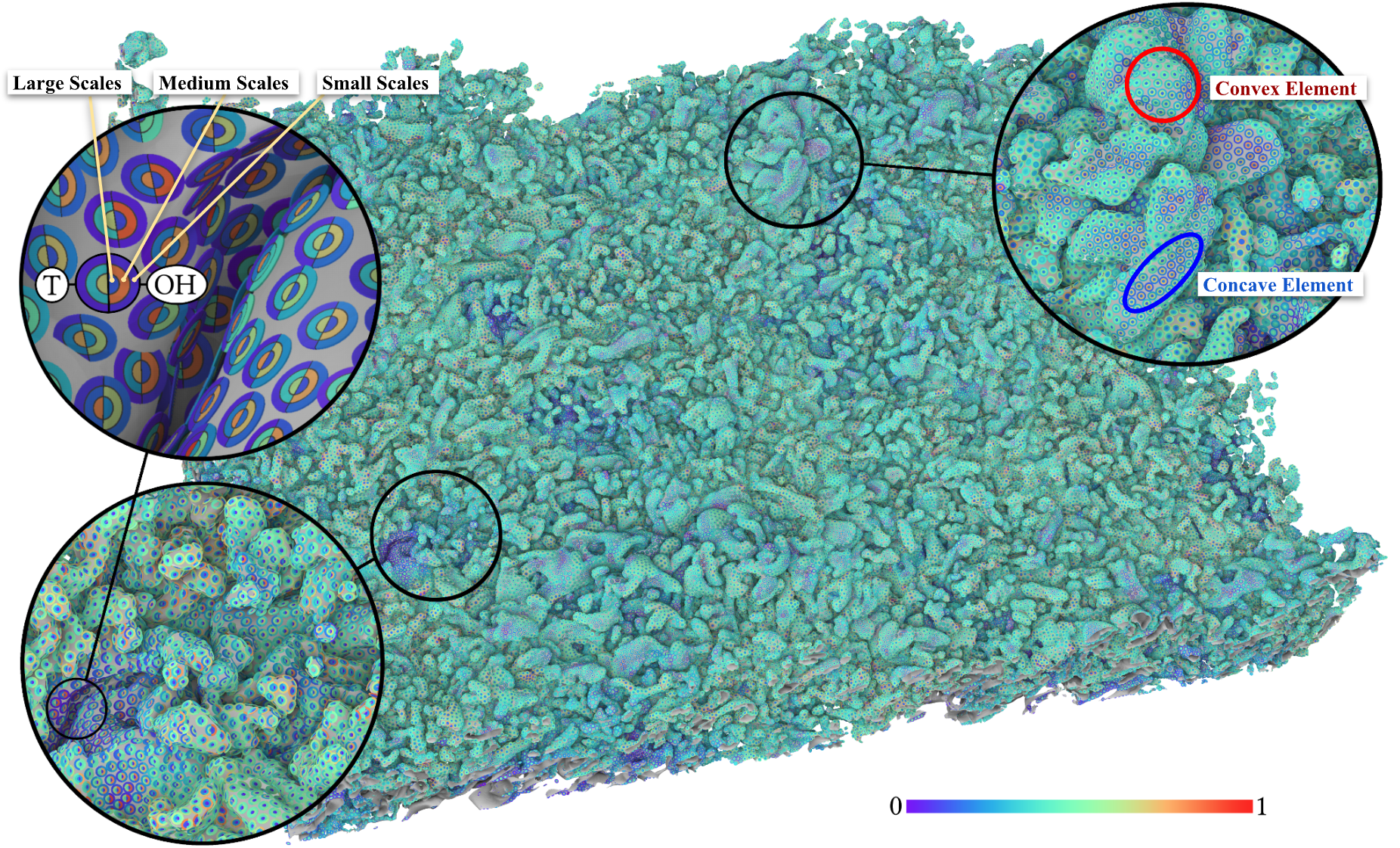}
      \caption{Radial glyph rendering with 3 scale bins for temperature in the combustion data. \color{changed} Right zoom: temperature glyphs tend to show different colors between the convex (circled in red) and concave (circled in blue) regions. \color{black} Left zoom: glyphs are configured to show temperature and hydroxyl radical (OH) together (left side and right side of glyph respectively).  In some areas, the balance of scale contributions is very similar between the fields, while in other areas we see a major difference, e.g., the purple area where the medium scales are very dominant for OH compared to temperature. The zoomed-in regions are sized to give a sense of what the visualization might look like fully zoomed out on a high-resolution monitor. While the visibility of the glyphs from afar can be limited depending on the coarseness and size of the glyphs, patterns can still be perceived throughout and zoomed in on to see more precise detail.}
      \label{fig:case_1_1}
      \vspace{\figvspace}
\end{figure*}

\begin{figure*}[ht]
    \centering
    \includegraphics[width=0.97\linewidth]{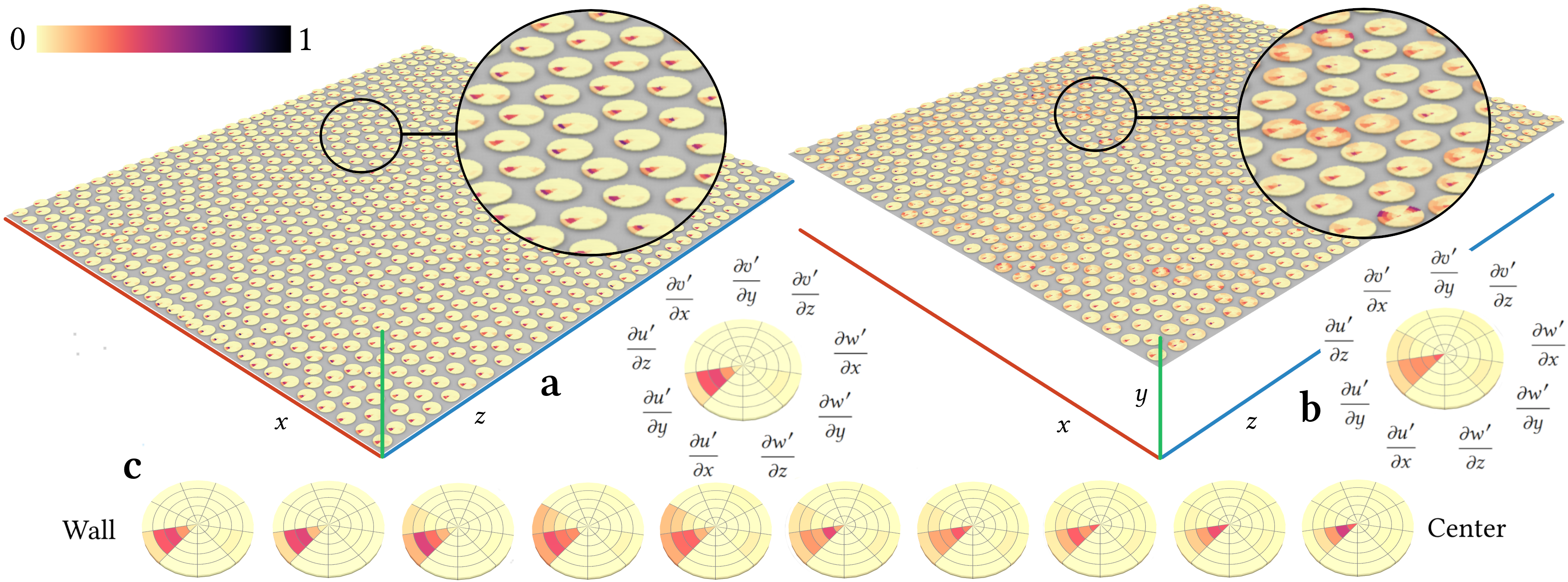}
    \caption{Radial glyphs showing %{\color{changed} 
    contributions of nine %} 
    components of the dissipation rate of turbulent kinetic energy over five scales in turbulent channel flows (x-axis: orange, y-axis: green, z-axis: blue).
    (a) an $x-z$ plane in the near-wall region. 
    (b) an $x-z$ plane in the outer region. 
    %Near to the wall, the energy dissipation is more strongly dominated by $(\partial u'/\partial y)^2$, and towards the center, while $(\partial u'/\partial y)^2$, contributions vary spatially. 
    (c) the average of each component over the $x-z$ plane from the wall (left) to the center (right). 
    %As you move from the wall to the center you see a shift in contribution towards the larger scales.
    }
    \label{fig:channel_case_1}
    \vspace{\figvspace}
\end{figure*}

\begin{table}[htb]
\caption{Storage for different dataset sizes comparing raw scale decomposition data and the aggregated statistical data used to generate the glyphs. The decomposition includes 4 bands, and the number of Voronoi regions/glyphs is given for all 4 bands.}
\centering
\setlength{\tabcolsep}{4pt} 
\small 
\begin{tabular}{@{} l r rr @{}} 
& \textbf{Raw Data} 
& \multicolumn{2}{c}{\textbf{Statistical Data}} \\
\cmidrule(lr){2-2} \cmidrule(lr){3-4}
\textbf{Grid Size} 
& \textbf{Per $X_{ij}$ (bytes)} 
& \textbf{Regions} & \textbf{Per $X_{ij}$ (bytes)} \\
\midrule
$\mathbf{233\times380 \times236}$ 
& 83,581,760 & 95,770 & 383,080 \\
$\mathbf{466 \times380 \times472}$ 
& 334,327,040 & 318,208 & 1,272,832\\
$\mathbf{699 \times 380 \times 708}$ 
& 752,235,840 & 820,323 & 3,281,292\\
$\mathbf{932 \times 380 \times 944}$ 
& 1,337,308,160 & 1,475,008 & 5,900,032 \\
$\mathbf{1165 \times 380 \times 1180}$ 
& 2,089,544,000 & 2,309,238 & 9,236,952\\
$\mathbf{1628 \times 380 \times 1180}$ 
& 2,919,980,800 & 3,230,126 & 12,920,504\\
% \bottomrule
\end{tabular}
\label{table:storage}
\vspace{-0.2in}  
\end{table}

\begin{table}[htb]
\caption{Render times over different data sizes for the strength glyph (STG), star glyph (SRG), and underlying isosurface (ISO). \textbf{Iso. Verts.} denotes the number of vertices in the isosurface mesh. As only one layer/band of glyphs are rendered at once, we report the number of glyphs in the single rendered band.}
\centering
\footnotesize
\setlength{\tabcolsep}{2.5pt}
\begin{tabular}{@{} l rrrrr @{}} % 
\textbf{Grid Size} & \textbf{Glyphs} & \textbf{Iso. Verts.} & \textbf{STG (s)} & \textbf{SRG (s)} & \textbf{ISO (s)} \\ % 
\midrule

$\mathbf{23\times380\times236}$ & 20,803 & 5,300,967 & 0.004 & 0.004 & 0.052 \\

$\mathbf{466\times380\times472}$ & 73,754 & 19,676,313 & 0.019 & 0.020 & 0.173 \\

$\mathbf{699\times380\times708}$ & 182,993 & 48,339,606 & 0.026 & 0.026 & 0.412 \\

$\mathbf{932\times380\times944}$ & 337,574 & 89,828,406 & 0.038 & 0.036 & 0.767 \\ 

$\mathbf{1165\times380\times1180}$ & 539,023 & 144,516,267 & 0.059 & 0.056  & 1.246 \\ 

$\mathbf{1628\times380\times1180}$ & 879,746 & 248,454,057 & 0.079 & 0.074 & 2.135 \\

% \bottomrule
\end{tabular}
\label{table:rendering}
\end{table}

% unrounded
% \begin{table}[htb]
% \centering
% \small
% \setlength{\tabcolsep}{4pt}
% \begin{tabular}{@{} l rrrrr @{}} % 
% \textbf{Grid Size} & \textbf{Glyphs} & \textbf{Iso. Verts.} & \textbf{STG (s)} & \textbf{SRG (s)} & \textbf{ISO (s)} \\ % 
% \midrule

% $\mathbf{233,380,236}$ & 20,803 & 5,300,967 & 0.0038 & 0.0041 & 0.0517 \\

% $\mathbf{466,380,472}$ & 73,754 & 19,676,313 & 0.0187 & 0.0196 & 0.1727 \\

% $\mathbf{699,380,708}$ & 182,993 & 48,339,606 & 0.0255 & 0.0261 & 0.4117 \\

% $\mathbf{932,380,944}$ & 337,574 & 89,828,406 & 0.0381 & 0.0363 & 0.7674 \\ 

% $\mathbf{1165,380,1180}$ & 539,023 & 144,516,267 & 0.0588 & 0.0556  & 1.2458 \\ 

% $\mathbf{1628,380,1180}$ & 879,746 & 248,454,057 & 0.0787 & 0.0744 & 2.1352 \\

% % \bottomrule
% \end{tabular}
% \caption{Render times over different data sizes for the strength glyph (STG), star glyph (SRG), and underlying isosurface (ISO). \textbf{Iso. Verts.} denotes the number of vertices in the isosurface mesh. As only one layer/band of glyphs are rendered at once, we report the number of glyphs in the single rendered band.}
% \label{table:rendering}
% \end{table}

\section{Performance, Scalability, and Data Reduction} %R4-C7
\label{sec:evaluation}

All measurements were performed on a single machine running Ubuntu 24.04, equipped with 32 GB of RAM, an Intel(R) Core(TM) i3-9100F CPU @ 3.60GHz, and a GeForce GTX TITAN X GPU with 12 GB of VRAM. We do all rendering tests on a 4k monitor, with the 3D view taking up $3072 \times 1990$ pixels. The rendering was done using an OpenGL 4.2 core profile context, with 16x multisampling (for anti-aliasing). 

\paragraph*{Preprocessing.} 
While we do not contribute novel algorithms for the preprocessing or claim the implementations we use are optimal, we present measurements of computation time to give a sense of the scalability. 
We break down the preprocessing into 4 stages, multiscale decomposition (MS), the distance field (DF), the LSRCVT, and multiscale aggregation (MA). 

We report performance for a small $256^3$ case (16,777,216 grid points), like what is depicted in Figure~\ref{fig:compare_glyphs}, as well as the larger case depicted in Figure~\ref{fig:case_1_1} of size $1628\times380\times1180$ (729,995,200) grid points. In each case, we chose the same glyph density as depicted in the corresponding figures, which is 0.001875, and 0.015 respectively. Note, since the latter data was downsampled by a factor of 2 in each dimension, and the glyph density parameter is with respect to voxels, 0.001875 and 0.015 represent the same glyph density per unit volume in this case. For each case, the tessellation includes bands, radiating outward from the flame surface at intervals of 4 voxels, and 2 voxels respectively (again due to downsampling of the latter data, these parameter values represent the same distances in physical units).

For the smaller data, we used mirror extension and for the larger data, we used the multiple overlapping blocks approach. The multiblock implementation is a hybrid distributed and out-of-core approach based on MPI. Each rank is assigned a share of the blocks, and loads, computes, and writes each block assigned to it one at a time. Likewise, the LSRCVT and distance function are computed similarly in blocks. This approach makes it possible to process datasets that would otherwise be too large for the system memory or GPU memory to accommodate. Since the original distance field code was sequential, the distances at the block boundaries may not be accurate. However, we ensured the distances were correct within the range used in the tessellation (defined by the distance bands in the tessellation configuration) by having ghost regions around each block larger than the width of the tessellation bands. In the future, a more scalable parallel distance field implementation would significantly improve preprocessing time. 

Since the Curvelab CTF implementation internally performs many DFTs on the CPU using FFTW, we also experimented by replacing the FFTW calls with a GPU-accelerated DFT backend using cuFFT, and report the performance of both. For the CTF and aggregation stages, we process each variable separately using an identical process, so we report the performance per variable. We used 3 scale bins (1 forward transform and 3 backward transforms). Since we measured the performance on one node with only one GPU, performance gains by using more ranks were limited. We used only 1 rank for the CTF (since beyond that we run out of memory), 3 ranks for the distance function, and 2 ranks for the LSRCVT. 

The compute times (hh:mm:ss) for the $256^3$ case, and the $1628\times380\times1180$ are (MS 5:53, MS CUDA 3:24, DF 1:08, LSRCVT 0:37, MA 0:02), and (MS 1:22:47, MS CUDA 43:22, DF 46:56, LSRCVT 15:43, MA 0:31) respectively. Using a CUDA backend for the DFTs in the CTF resulted in almost a 2x speedup for MS.

\paragraph*{Rendering.} Since the visualization includes both the glyphs themselves and an isosurface beneath it, we report the rendering times for each rendering task separately. We tested the rendering performance using portions of the combustion dataset depicted in Figure~\ref{fig:case_1_1}. The glyph density parameter was $0.015$, as was also used in the rendering in Figure~\ref{fig:case_study_1_2}. This is a relatively high glyph density. Table~\ref{table:rendering} shows the results. As the isosurface has many more total triangles than the set of glyphs, the rendering for the glyphs ranges from 10 to 30 times faster. The bottleneck to rendering scalability is the isosurface rendering, which could be alleviated by rendering without the underlying isosurface if absolutely necessary, or by simplifying the mesh to reduce the triangle count.
 
\paragraph*{Data reduction.} The scale decomposition produces one new volume for each scale bin and field. However, only the spatially aggregated statistics, glyph positions, and normals are needed to produce our glyph-based visualizations. In Table~\ref{table:storage}, we show the storage sizes for the much-reduced glyph data in comparison with the raw scale decomposition data.\color{black}

\section{Case Studies}
\label{sec:case-studies} 

To demonstrate the use of our system, we conducted two case studies visualizing compressible combustion flow and incompressible channel flow data generated by turbulence simulations. The former examines the behavior of gaseous species with changing volume. The latter examines the behavior of constant-density fluid in an enclosed space. \color{changed} These studies were conducted in collaboration with three domain experts, 
%: two in turbulent combustion, and one in channel flow. The experts 
who selected the datasets and parameters for preprocessing beforehand, as described in Sections~\ref{ctfmd} through~\ref{sec:aggregation}, then coordinated with us to preconfigure 
%their 
visualization system, as described in Section~\ref{sec:vis-system}, at the time of their respective %case 
studies.\color{black}

\subsection{Turbulent Combustion}

The dataset chosen for this case study is a fuel-lean ammonia/hydrogen/nitrogen-air turbulent flame at elevated pressure~\cite{rieth2022enhanced}, representing conditions in a gas turbine combustor burning carbon-free fuel. While blends of ammonia and hydrogen present a promising carbon-free fuel, there are challenges associated with so-called thermo-diffusive instabilities, which form due to the fast diffusion of hydrogen, amplifying flame propagation speeds in regions that are convex towards the unburned mixture and diminishing flame speeds in concave regions. \color{changed} The flame front, a sheet-like region where heat release occurs, is separated by reactants on one side and products of combustion on the other side. Through turbulent strain and the flame's self-propagation, the flame front is geometrically distorted and wrinkled forming local minima, maxima, ridges and valleys. In particular, hydrogen preferentially diffuses towards, or focuses, where the front's local curvature is convex towards the unburnt reactants relative to heat and other species. This leads to local enrichment of the fuel and a higher equivalence ratio. Conversely, hydrogen is defocused when the curvature is concave towards the reactants~\cite{rieth2022enhanced, SHI2023100105}. Local hydrogen enrichment leads to higher heat release, burning rates and temperatures that affect emissions of nitric oxide and nitrous oxide. \color{black} Such instabilities form intrinsic length scales, \color{changed}corresponding to a flame thickness, but also interact with broadband turbulence. \color{black}Understanding the complex interplay between thermo-diffusive instabilities that manifest at flame scales and turbulence across many decades of length scales is crucial for the development of next-generation clean and carbon-free engines utilizing hydrogen-rich fuels.    

Figure~\ref{fig:case_1_1} shows a multiscale glyph visualization of the temperature field for a single snapshot. Figure~\ref{fig:case_study_1_2} presents the glyph-based decomposition over a $256^3$ region in more detail for temperature, hydroxyl radical (OH), and hydrogen peroxide (H$_\mathrm{2}$O$_\mathrm{2}$), which are both important species formed in the combustion process. The hydroxyl radical forms at high temperatures, while hydrogen peroxide is mostly present at intermediate temperatures. Glyphs are computed on the band closest to the flame. Here, the surface representing the flame front is defined based on a temperature isovalue of 1700 K, which is close to the maximum flame temperature of an equivalent unperturbed flame, allowing for the % R4-C19
analysis of the complex turbulent flame front. The tool allows for direct visualization of the scale dependence of relevant quantities, thus helping to elucidate the complex physics present in turbulent flames subject to thermo-diffusive instabilities. The visualization clearly highlights the differences between convex and concave flame elements \color{changed}(see Figures~\ref{fig:case_study_1_2} and~\ref{fig:case_1_1})\color{black}. In previous studies of \color{changed} preferential diffusion of hydrogen-rich fuels (including conventional analysis of this dataset), conditional statistics in composition space (i.e. conditional on temperature, a variable denoting the progress of reactions in a premixed flame)  have shown that a flame's local topology, i.e. the local curvature of the flame front is correlated with the burning intensity and species gradients which lead to local hydrogen enrichment in sections of the flame front that are curved convex towards the reactants and vice-versa for concave curved sections~\cite{rieth2022enhanced}. \color{black}{With the visualization presented here, for the first time,} \color{changed} these physical insights are readily gleaned from \color{black}visualization of the flame in physical space where scale dependent statistics can be directly correlated with the local flame topology. \color{black}{The visualization clearly shows that concave }\color{changed}sections of the flame front \color{black}are mostly dominated by large-scale turbulence features (i.e., turbulent mixing \color{changed}from large eddies \color{black}rather than flame propagation), whereas convex \color{changed}sections of the flame burning with high intensity are controlled \color{black}by smaller scales closer to (or smaller than) the flame scale. Moreover, it can be observed that temperature overall is dominated by scales larger than the hydroxyl radical. Figure~\ref{fig:case_study_1_2} further combines scale information in physical and composition space (the space spanned by the thermochemical scalars), i.e. relating chemical species and temperature. 
While chemical species concentrations and temperature are affected by transport through the underlying flow field as well as local chemical reactions, there are significant differences in how these quantities evolve, and how they are affected by local variations in the thermochemical state, as is demonstrated by the difference in their scale-dependence. In the scatter plot view, regions of high temperature and significant contribution from the large scales are selected. In the 3D view, these regions correspond to a convex \color{changed}section of the flame\color{black}. With the linked view, regions in physical and composition space and their scale dependence can directly be related. \color{black}This shows that concave \color{changed}sections of the flame\color{black}, which are dominated by large-scale features, reside in regions of \color{changed}low \color{black}intermediate hydroxyl mass fractions, indicating that these burn less intensely compared to other convex \color{changed}{flame sections}. \color{black} Conversely, but not shown for brevity, a selection of regions that are affected by flame scales shows \color{changed}a prevalence for \color{black}these convex flame features. \color{black}Overall, this visualization demonstrates how \color{changed}the local flame geometry exhibits \color{black}different scale dependence, providing valuable insight into the behavior of hydrogen-enriched flames. \color{changed}This visualization approach enables scientists to understand complex multi-scale, multi-variate information and its correlation with the local physical topology of both the flames and the flow (not shown here) encapsulated in a single visualization. By comparison, traditional analysis approaches require multiple visualizations in physical and composition space combined with multi-scale statistical analysis to arrive at a summary of the relevant correlations.\color{black}

\begin{figure} %R4-C13
     \centering
     \includegraphics[width=0.98\linewidth]{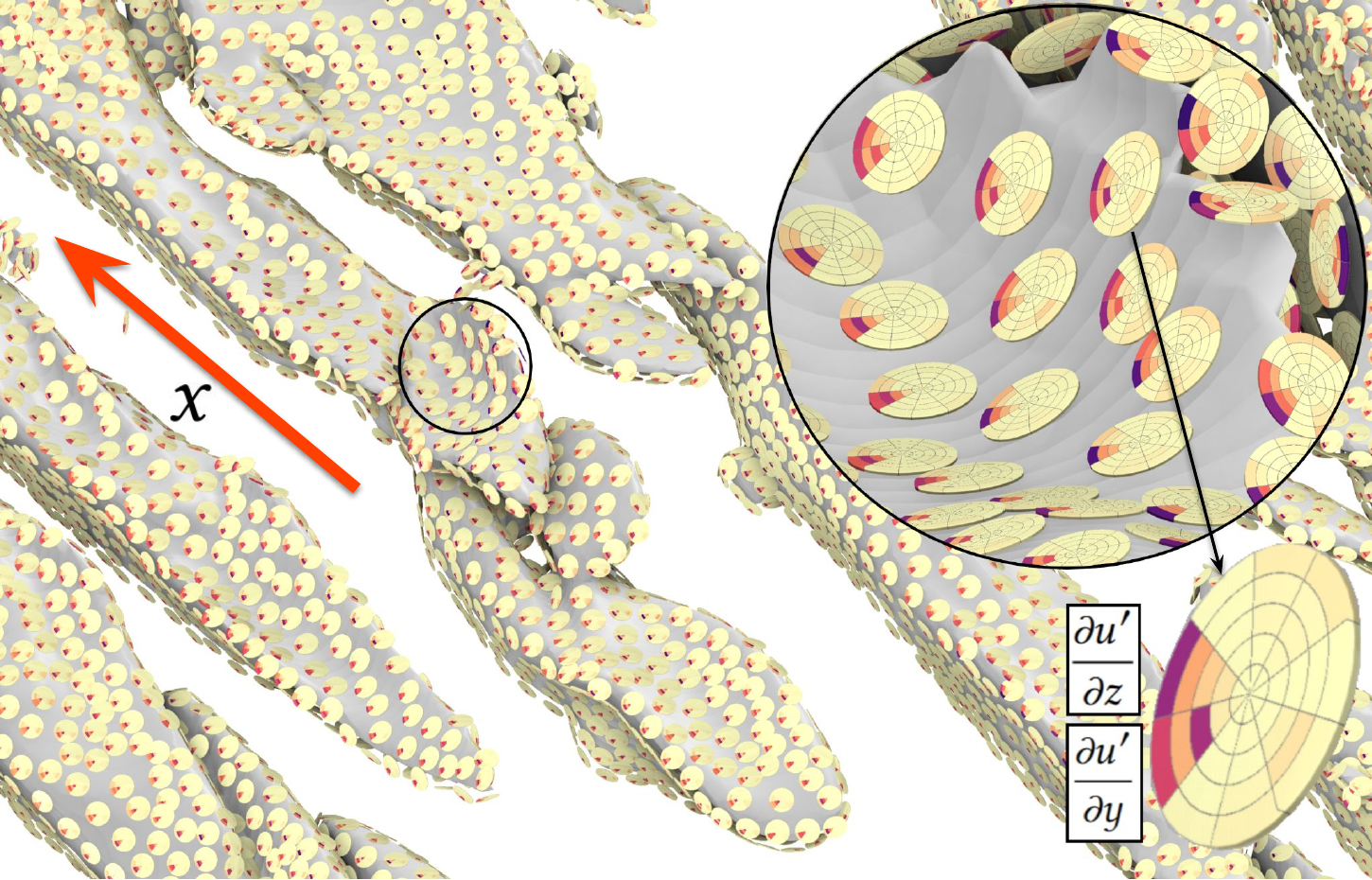}
     \caption{Radial glyphs showing components of the dissipation rate of turbulent kinetic energy over five scales in turbulent channel flows. The glyphs are placed over an isosurface, with isovalue 150.0, in the dissipation rate of the turbulent kinetic energy in the near-wall region. The highlighted wedges consistent across glyphs show that energy dissipation is dominated by the corresponding $\partial u'/\partial y$ component. However, especially where neighboring structures come close together, we see higher non-$\partial u'/\partial y$ contributions. One such region is zoomed in to show glyphs whose non-$\partial u'/\partial y$ contributions, mainly $\partial u'/\partial z$, are influenced by the neighboring structures.}
     \label{fig:channel_case_2}
      \vspace{\figvspace}   
\end{figure}

\subsection{Turbulent Incompressible Channel Flow}
%Figure xxx shows the decomposition of the scale based on glyphs of the pressure-strain components of the turbulent kinetic energy (TKE). TKE is calculated from the variances of each velocity component:
%\[\text{TKE} = \frac{1}{2} \left( \langle u'^2 \rangle + \langle v'^2 \rangle + \langle w'^2 \rangle \right),\]
%where  $u$,  $v$, and $w$  denote the velocity components in each spatial direction,  $'$  indicates fluctuations from the mean, and  $\langle \cdot \rangle$  represents a spatio-temporal average.
%In turbulent channel flow, only  $\langle u'^2 \rangle$  has a non-zero production term. Therefore, it is important to understand how the energy generated in  $\langle u'^2 \rangle$  is transferred to the other components. The pressure-strain terms for each component,  $\langle p' (\partial u'_\alpha/\partial x_\alpha) \rangle $, serve as the mechanism that transfers energy between components. Note that
%\[
%p'\frac{\partial u'}{\partial x} +
%p'\frac{\partial v'}{\partial y} + 
%p'\frac{\partial w'}{\partial z} = 0
%\]
%due to the continuity constraint.
%Despite its importance, understanding the interaction between pressure-strain %terms and the coherent structures of turbulent flows remains incomplete.

Figure~\ref{fig:channel_case_1} illustrates the decomposition of the scale based on the glyphs of the dissipation rate of the turbulent kinetic energy, $\epsilon_K$, in incompressible turbulent channel flows using the data from DNS of turbulent channel flows at friction Reynolds number 182 \cite{Lee.2015wa}. Here, $\epsilon_K = \mu\sum_{i,j} (\partial u_i' / \partial x_j)^2/2$ where $u_i'$ is velocity fluctuations in streamwise ($x$), wall-normal ($y$), and spanwise ($z$) directions. Also, $\mu$ is the dynamic viscosity of a fluid. Note that there are nine components of $\epsilon_K$. Understanding the behavior of $\epsilon_K$ is critical for analyzing the lifecycle of turbulent kinetic energy, particularly in the development of subgrid-scale (SGS) models for large-eddy simulations. 
While each component of $\epsilon_K$ is often modeled as being isotropic, different components exhibit varying contributions depending on the region within the flow domain. This variability makes it challenging to construct universal SGS models. Unlike homogeneous isotropic turbulence, turbulent channel flows exhibit strong anisotropic behavior, especially in the near-wall region, where all flow quantities display distinct characteristics.
Figure~\ref{fig:channel_case_1}a shows the glyphs of $\epsilon_K$ in the near-wall region. In this region, component $\partial u'/\partial y$ dominates over others. Furthermore, the dominant length scales of $\partial u'/\partial y$ are small.  
In contrast, the glyphs of $\epsilon_K$ in the channel center region exhibit different behavior, as shown in Figure~\ref{fig:channel_case_1}b. Compared to the near-wall region, the flow in the channel center displays more contributions from components other than $\partial u'/\partial y$. Also, noticeably, the length scale of each component is larger than the dominant length scale in the near-wall region as shown in Figure~\ref{fig:channel_case_1}c. This is consistent with the theory that the dissipation length scale, $\lambda_\epsilon$, grows with the wall-normal distance, $y$, as $\lambda_\epsilon \sim y^{1/4}$\cite{Lee.20196fe}. 
%Notably, the contributions of the longitudinal components $(\partial u / \partial x)^2$, $(\partial v / \partial y)^2$, and $(\partial w / \partial z)^2$ are approximately equal. A similar trend is observed among the transverse components, such as $(\partial u / \partial y)^2$, $(\partial u / \partial z)^2$, $(\partial v / \partial x)^2$, etc.
%Furthermore, the results show that the contribution of the transverse components is approximately twice that of the longitudinal components, a behavior commonly referred to as isotropic dissipation (CITE POPE). 
%Additionally, the dominant length scales of each component in the channel center region are generally larger than those in the near-wall region. 

Additionally, Figure~\ref{fig:channel_case_2} illustrates the relationship between the coherent structures of $\epsilon_K$ and the contribution of each component. In the near-wall region, the dissipation rate exhibits structures that are elongated in the $x$-direction. As previously mentioned, the dominant contribution originates from the $\partial u'/\partial y$ component. However, when the two structures are in close proximity, the glyphs indicate significant contributions from other components. Understanding the interaction between coherent structures is crucial for unraveling the self-sustaining cycle of turbulence \cite{Jimenez.1999}, and this feature can be applied to various aspects of turbulent flow research. \color{changed}The structures highlighted in Figure~\ref{fig:channel_case_2} are primarily located at $y\approx 0.05$–$0.2\delta$, where near-wall dynamics are particularly energetic. Here, $\delta$ denotes the channel half-width. A strong contribution from $\partial u’/\partial y$ is expected due to wall-induced gradients and anisotropy. However, during coherent structure interactions, $\partial u’/\partial z$ becomes non-negligible (e.g., Figure~\ref{fig:channel_case_1}c and Figure~\ref{fig:channel_case_2}), suggesting that $\partial u’/\partial z$ may serve as a useful indicator of connectivity between coherent structures and dissipation statistics. In contrast, $\partial u’/\partial x$ remains comparatively small in these cases. Finally, these results corroborate the growth of the dissipation length scale with $y$, underscoring the potential of glyph-based multiscale analysis for high-$Re$ flows, where pronounced scale separation and cross-scale interactions remain open questions in wall-bounded turbulence \cite{Chen.2021,Jimenez.2024,Lee.2025}.\color{black}

Overall, the glyph-based visualization enables component-resolved interrogation of anisotropic, multiscale structures in incompressible wall-bounded turbulence with strong shear, and offers a compact framework for connecting coherent structures, dissipation physics, and modeling needs. \color{changed} Extensive prior work, including well-established criteria such as $\lambda_2$ method~\cite{Jeong_Hussain_1995} along with subsequent studies applying these techniques, on identifying vortex and coherent structures in (wall-bounded) turbulent flows has been done. However, applying such methods to perform systematic and statistically meaningful analysis in high-Reynolds-number turbulent flows remains challenging for several reasons. One is that the inherently chaotic nature of turbulence and the large number of vortical structures at high Reynolds numbers make comprehensive identification and tracking difficult. Another is that linking individual coherent structures to global statistical properties of turbulence typically requires extensive post-processing, which limits scalability and general applicability. As a result, many existing coherent-structure identification methods have primarily been used as proof-of-concept or qualitative visualization tools rather than as practical frameworks for quantitative, large-scale analysis. In contrast, the present approach is designed to address these limitations. This enables a more systematic connection between individual coherent structures and the statistical description of turbulent flows. \color{black}

%In summary, this glyph-based visualization facilitates the analysis of anisotropic structures with multiscale behavior in incompressible turbulent flows characterized by strong shear stress. 

%There is extensive prior work on identifying vortex and coherent structures in (wall-bounded) turbulent flows, including well-established criteria such as \lambda_2 method (e.g., Jeong & Hussain, J. Fluid Mech., 1995), and thousands of subsequent studies applying these techniques. However, applying such methods to perform systematic and statistically meaningful analysis in high-Reynolds-number turbulent flows has remained challenging for several reasons. First, the inherently chaotic nature of turbulence and the large number of vortical structures at high Reynolds numbers make comprehensive identification and tracking difficult. Second, linking individual coherent structures to global statistical properties of turbulence typically requires extensive post-processing, which limits scalability and general applicability. As a result, many existing coherent-structure identification methods have primarily been used as proof-of-concept or qualitative visualization tools rather than as practical frameworks for quantitative, large-scale analysis. In contrast, the present approach is designed to address these limitations. This enables a more systematic connection between individual coherent structures and the statistical description of turbulent flows.

\section{Discussion}
\label{sec:discussion} 
Based on discussions with domain experts and the case study results, the strength glyph was the most favored of the glyph designs. This is likely because of its ability to display contribution across fields and scales in an easily perceived and controlled way, making it most conducive to multiscale multi-physics analysis at different viewing distances. Following this is the starplot glyph, which provides an alternative way to compare contributions across fields for a fixed scale that inherits the perceptual benefits of spatial visual encoding and may be especially useful to those who struggle with color-based visual mapping. While the same information could be visualized with multiple views without multivariate glyphs, encoding one scale and variable per view results in a large amount of perceptual overhead in the need to look across different views while keeping mental track of spatial locations and values. Additionally, multiplying the number of views impacts performance, memory use, and screen space. 

\subsection{Limitations} %R4-C18
Given the nature of our visualization, the most evident limitation comes from placing glyphs along small, highly concave, or folded surfaces, as such deformations can hide them. Glyphs might also be occluded should they intersect surfaces with narrow folds. While our case studies observe ideal usage along convex surfaces for select regions of interest, other noisier datasets may suffer from surface-based occlusion. \color{black}  One way of mitigating this may include applying a flattening deformation on a select region of the surface, either directly on the volume view or in a separate view, so that all glyphs in that region can be seen without deformation or occlusion. This, of course, would come at the cost of seeing the 3D geometry of the surface beneath it. Another approach may be to isolate a specific region of the surface while removing the rest from view. While this could help reduce occlusion resulting from overlapping surfaces, the glyphs are still subject to distortion based on their orientation relative to the viewport.

Limitations also arise as a result of how the discrete curvelet transform works and is implemented, \color{changed}such as \color{black}the assumptions of periodic boundary conditions and uniform grid spacing. If these requirements are not satisfied, it can introduce unwanted boundary artifacts and other distortions in the final output. While we apply preprocessing on the incoming data collaboratively with domain scientists to produce sound results, it can generally be a non-trivial task to meet these requirements \color{changed} in a way that \color{black} avoids introducing artifacts, distorting the data, or ending up with data that is too large to process once resampled or padded. This is especially the case when dealing with non-periodic boundaries, and complex or highly anisotropic grid configurations.
%, and expert knowledge of the simulation and science domain may be needed to process the data appropriately on a case by case basis. (R4-C9) 
In particular, clipping cuts structures, introducing sharp edges that didn't exist before. This requires subsequent mirror extension or windowing to resolve additional artifacts. For scalability, we rely on the overlapping windowed blocks approach. However, processing the whole domain at once would give a broader, more detailed picture of the large-scale contributions. These have the most direct impact on the quality and scalability of the multiscale decomposition. 

%(R4-C8)
As pointed out in~\ref{sec:tessellation}, there are limitations on the granularity of information captured and glyph resolution with respect to site density. Higher density entails more sites to capture information at a higher fidelity but reduces glyph resolution, as glyphs have to be shrunk to accommodate an increased site count. Conversely, a lower density entails coarser granularity, enhancing glyph visibility at the cost of capturing variation. This could be mediated by computing multiple tessellations with increasing site densities then applying semantic zooming to change the tessellation and its corresponding glyph visualization. Magic lenses~\cite{10.1145/3596711.3596719, 10.1145/274644.274714} could support localized interactive zoom and detail level within a global context. For especially large volumes, surfaces could be color encoded to show variation or quantities of interest across fields and scales while zoomed out, and the glyphs could be revealed upon zooming in.\color{black}  

%i.e. there will be more glyphs along the surface to show the variation across regions, 
\subsection{Additional Considerations} %(R4-C6)
We originally used the curvelet transform for its geometry-preserving properties as utilized in~\cite{BERMEJO-MORENO_PULLIN_2008}. While we did not incorporate geometric features into our glyphs, this work could be used as a starting point in developing glyphs or similar representations for that purpose. Future work could also focus on experimenting with and validating different multiscale decomposition approaches in various applications, as well as their scalability. \color{black} This work could also be extended such that multiscale features can be compared across different time steps as well as spatial regions. Our approach might also benefit from having additional integrated statistical visualizations similar to the work of Abbasloo et al.~\cite{7192624}, where they apply glyphs and other visual representations of statistical values from tensor fields, like the mean or variance, to cross sections of a corresponding volume. This encourages additional exploration of said volume while mitigating occlusion brought on by placing glyphs along its surface. We could also make more explicit use of topology in our visualization by incorporating graph representations of aggregate regions, especially in conjunction with the glyphs. 
\section{Conclusion}
\label{sec:conclusion} %(R1-C3)
Our glyph-based visualizations applied to the analysis of turbulent combustion show valuable insight into the behavior of hydrogen-enriched flames. For the first time, the presented visualizations directly link differences in scale dependence to flame topologies and composition space. This is crucial for understanding the interaction of thermo-diffusive instabilities occurring in hydrogen-enriched flames necessary to develop predictive combustion models for the design of engines utilizing carbon-free fuels. It also demonstrates that a glyph-based system can provide valuable insights into the multiscale anisotropic behavior of turbulent flows with strong shear stresses, such as those observed in turbulent channel flows. Given this, our system demonstrates new capabilities that can enhance analysis of scale dependence within a volumetric multi-physics context, especially if it can be made more scalable and incorporate other types of analyses.

\bibliographystyle{IEEEtran}
\bibliography{egbibsample.bib}       

\end{document}